\documentclass[runningheads]{llncs}

\usepackage[final]{eccv}

\usepackage{eccvabbrv}

\usepackage{microtype}
\usepackage{graphicx}
\usepackage{subcaption}
\usepackage{booktabs} %

\usepackage[accsupp]{axessibility}  %

\usepackage{hyperref}

\usepackage{orcidlink}

\usepackage{amsmath}
\usepackage{amssymb}
\usepackage{mathtools}

\usepackage{graphicx}
\usepackage{booktabs}
\usepackage{multicol}

\usepackage[capitalize,noabbrev]{cleveref}

\usepackage{url}
\usepackage{adjustbox}
\usepackage{epsfig}
\usepackage{graphicx}
\usepackage{amsmath}
\usepackage{amssymb}
\usepackage{wrapfig} 
\usepackage{xcolor}
\usepackage{subfloat}
\usepackage{multirow}
\usepackage{booktabs}
\usepackage{makecell}
\usepackage{mdframed}
\usepackage{colortbl}
\usepackage{caption}

\usepackage{xspace}
\def\onedot{.\xspace}

\def\eg{\emph{e.g}\onedot}

\def\vs{\emph{vs}\onedot}

\def\Model{\textbf{OpenVision 3}}

\usepackage[textsize=tiny]{todonotes}

\title{OpenVision 3: A Family of Unified Visual Encoder for Both Understanding and Generation}
\titlerunning{Abbreviated paper title}

\begin{document}

\author{Letian Zhang\inst{1}\thanks{Equal contribution} \and
Sucheng Ren\inst{2}\protect\footnotemark[1] \and
Yanqing Liu\inst{1} \and
Xianhang Li\inst{1} \and
Zeyu Wang\inst{1} \and
Yuyin Zhou\inst{1} \and
Huaxiu Yao\inst{3} \and
Zeyu Zheng\inst{4} \and
Weili Nie\inst{5} \and
Guilin Liu\inst{5} \and
Zhiding Yu\inst{5} \and
Cihang Xie\inst{1}
}

\authorrunning{L. Zhang, S. Ren et al.}

\institute{
$^1$ UC Santa Cruz \quad $^2$ JHU \quad $^3$ UNC-Chapel Hill \\
$^4$ UC Berkeley \quad $^5$ NVIDIA
}

\maketitle

\begin{center}
\small 
   {\includegraphics[height=1.1em]{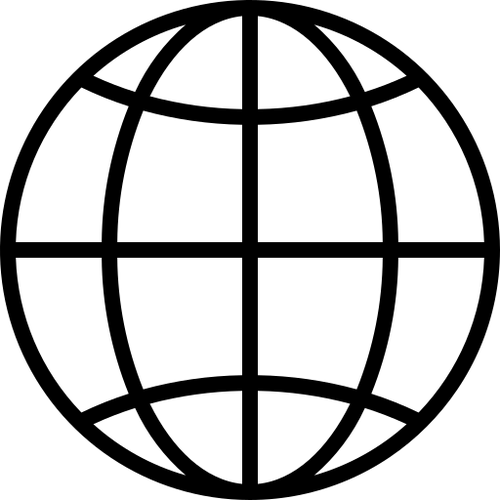}} \textbf{Project Page}: \url{https://ucsc-vlaa.github.io/OpenVision3/}
\end{center}

  \begin{center}
    \centering
    {\includegraphics[width=1\textwidth]{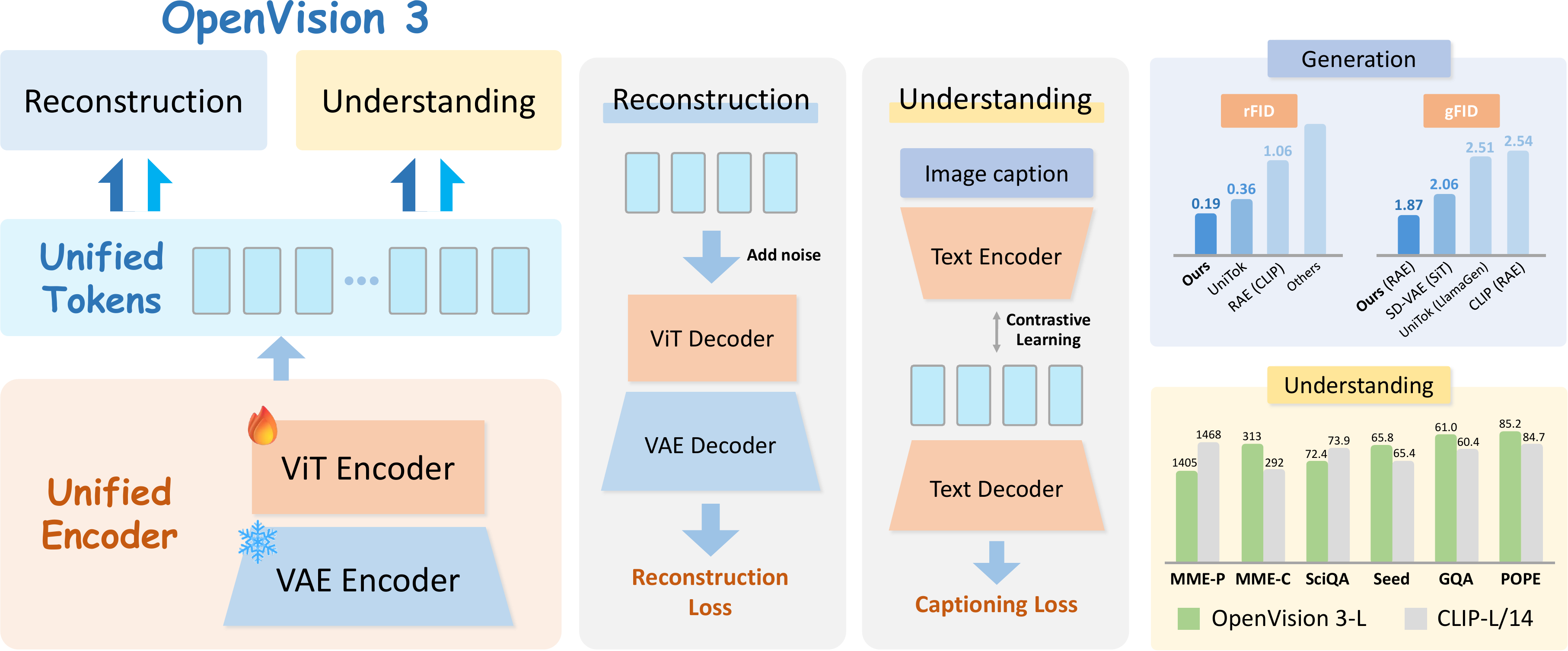}}
    \captionof{figure}
    {An overview of \textbf{\Model's architecture design and performance highlight}. 
    \textbf{Left panel}: The architecture of~\Model. We employ a frozen VAE and a trainable ViT as the unified tokenizer, which produces tokens that are fed simultaneously into both the reconstruction and understanding branches.
    \textbf{Middle panel}: The learning objectives of the two branches. For the reconstruction branch, we focus on pixel-level image reconstruction; concurrently, the understanding branch is optimized via joint contrastive and captioning objectives.
    \textbf{Right panel}: The performance summarization. The result shows that \Model~outperforms other unified tokenizers and semantics-based encoders in reconstruction and generation, while remaining competitive with CLIP in multimodal understanding ability. 
    }
    \label{fig:teaser}

  \end{center}

\begin{abstract}
This paper presents a family of advanced vision encoder, named \Model, that learns a \textit{single}, \textit{unified} visual representation that can serve both image understanding and image generation. Our core architecture is simple: we feed VAE-compressed image latents to a ViT encoder and train its output to support two complementary roles. First, the encoder output is passed to the ViT-VAE decoder to reconstruct the original image, encouraging the representation to capture generative structure. Second, the same representation is optimized with contrastive learning and image-captioning objectives, strengthening semantic features. By jointly optimizing reconstruction- and semantics-driven signals in a shared latent space, the encoder learns representations that \textit{synergize} and \textit{generalize} well across both regimes. We validate this unified design through extensive downstream evaluations with the encoder \textbf{\textit{frozen}}.
For generation, we test it under the RAE framework: ours substantially surpasses the standard CLIP-based encoder (\eg, gFID: 1.87~\vs~2.54 on ImageNet). 
For multimodal understanding, we plug the encoder into the LLaVA-1.5 and LLaVA-NeXT framework: it performs comparably with a standard CLIP vision encoder (\eg, 63.3~\vs~61.2 on SeedBench, and 59.2~\vs~58.1 on GQA).
We provide empirical evidence that generation and understanding are mutually beneficial in our architecture, while further underscoring the critical role of the VAE latent space.
We hope this work can spur future research on unified modeling.

\end{abstract}

\section{Introduction}
\label{sec:intro}

Unified Multimodal Models (UMMs) has emerged as a cornerstone of multimodal research, driven by the need for systems that seamlessly integrate visual understanding and generation. Their development is grounded in the Platonic Representation Hypothesis~\cite{huh2024platonic}, which posits that different data modalities reflect a shared underlying reality, and that learning a unified multimodal representation enables mutual benefits across modalities while improving generalization. The success of representative proprietary UMMs such as GPT-4o~\cite{hurst2024gpt}~and Gemini-2.5 Flash~\cite{comanici2025gemini}, as well as public models like BAGEL~\cite{deng2025emerging}, further supports this view, showcasing strong capabilities in dialogue-based multi-turn generation, multimodal in-context learning, and fine-grained control over generated content.

A key challenge in developing native UMMs lies in how visual representations are encoded. Owing to the representational discrepancy between visual understanding and visual generation, a common UMM design, exemplified by UniFluid~\cite{fan2025unified}, BAGEL~\cite{deng2025emerging}, and MOGAO~\cite{liao2025mogao}, employs two distinct visual tokenizers that encode the same image twice, producing one set of high-level semantic tokens and another set of low-level, pixel-reconstructable tokens. While effective, this approach increases system complexity and may hinder deeper synergy between understanding and generation.

Another line of work attempts to bridge this gap through shared visual tokenizers. However, these approaches typically rely on quantized hidden representations, which inevitably introduce discretization errors and limit generation quality (\eg, TokenFlow~\cite{qu2025tokenflow}, UniTok~\cite{ma2025unitok}, and EMU3.5~\cite{cui2025emu3}). As a result, developing a simple yet effective \textit{continuous} visual tokenizer that naturally supports both visual understanding and generation remains an open and practically challenge.

This paper presents \Model~as a step toward mitigating this challenge. Concretely, we build our tokenizer by stacking a ViT encoder on top of a well-trained VAE encoder. The output of the ViT encoder is further fed into two separate branches, one generation decoder that is trained to reconstruct the original image and enforce preservation of low-level visual information, and another understanding decoder that is trained by contrastive and captioning objectives, enhancing semantic supervision. Intriguingly, as analyzed in~\cref{sec:loss_interaction}, this design choice can non-trivially  \textit{synergize} the learning of both fine-grained details and high-level semantics, \eg, even optimizing understanding loss alone can lead to better reconstruction performance, and conversely, optimizing reconstruction alone can benefit semantic alignment. This behavior is also consistent with recent evidence that semantically informed tokenization can facilitate low-level reconstruction learning~\cite{yu2024representation, leng2025repa, yao2025reconstruction}, and may even serve as a direct drop-in replacement for purely reconstruction-oriented tokenizers~\cite{zheng2025diffusion}. 

Our experiments validate the effectiveness of \Model~across understanding, reconstruction and generation. Crucially, in all downstream evaluations we keep the tokenizer/encoder \textbf{frozen}, ensuring that the reported gains reflect the quality and transferability of the learned visual representation rather than task-specific fine-tuning.
For understanding evaluation, we integrate our tokenizer into the LLaVA-1.5 and LLaVA-NeXT framework for training and evaluate its performance across various standard multimodal benchmarks. For the reconstruction evaluation, we evaluate the quality of reconstructed images~\Model~on COCO~\cite{lin2014microsoft}~and ImageNet~\cite{deng2009imagenet}. For the generation evaluation, we train flow matching model following RAE~\cite{zheng2025diffusion} on ImageNet. The results demonstrate that~\Model~is comparable to CLIP in terms of understanding capabilities(\eg, 63.3~\vs~61.2 on SeedBench, and 59.2~\vs~58.1 on GQA), while surpassing existing unified tokenizers in image reconstruction (\eg, rFID: 0.187~\vs~0.362 on ImageNet). For image generation, our tokenizer outperforms standard CLIP-based encoder under the RAE framework by a large margin (\eg, gFID: 1.87~\vs~2.54 on ImageNet). 
Moreover, Through the ablation study, we observed that our tokenizer achieves a mutual promotion between understanding and generation during training. The utilization of the VAE latent space has also been empirically validated as both effective and indispensable through our experiments. We hope that releasing~\Model~will catalyze further research into more advanced unified vision tokenizers.

\section{Related Work}
\subsection{Vision-Language Pretraining}
Vision-Language pretraining serves as the cornerstone of multimodal representation learning. 
Pioneering works, exemplified by CLIP, adopt contrastive learning as their core methodology to extract and align visual and textual features. 
This training paradigm was subsequently adopted by a wide range of studies, such as LAION~\cite{schuhmann2022laion}, DataComp~\cite{gadre2023datacomp}, DFN~\cite{fang2023data}, OpenCLIP~\cite{cherti2023reproducible}, MetaCLIP~\cite{xu2023demystifying, chuang2025meta}~and CLIPA~\cite{li2023inverse, li2023clipa}. 
These research focuses primarily on efficient and open-sourced data and scaling methodologies.
Follow-up works have continuously explored alternative training regimes. 
CoCa~\cite{yu2022coca}~adds a captioning loss on the multimodal decoder outputs which predicts text tokens autoregressively. 
SigLip~\cite{zhai2023sigmoid}~proposes to replace contrastive loss with pairwise Sigmoid loss. 
SigLip2~\cite{tschannen2025siglip}~further extends this by incorporating captioning-based pretraining, self-distillation and masked prediction. 
The AM-RADIO~\cite{ranzinger2024radio, heinrich2025radiov2} series of works is dedicated to multi-resolution training and knowledge distillation from multiple teacher models. 
More recently, CLIPS~\cite{liu2024clips}, OpenVision~\cite{li2025openvision}, and OpenVision 2~\cite{liu2025openvision}~have focused on the efficient utilization of captioning loss in vision-language pretraining, demonstrating it to be a low-cost yet high-performance approach.
Our work builds upon this line of research and extend this efficient paradigm to unified multimodal learning. 
By combining contrastive, captioning, and reconstruction losses, we simultaneously supervise semantic and generative learning, resulting in reciprocal performance gains for both parts.

\subsection{Unified Tokenizer}

Extracting representative feature for both generation and understanding has been a bottleneck for the development of unified modeling. Previous works mostly adapt separate encoder for the two kinds of features, and then combine them. For example, BAGEL takes FLUX-VAE~\cite{flux2024, labs2025flux}~for low-level features and SigLIP2 for semantic features. UniWorld-V1~\cite{lin2025uniworld}~also computes the two types of features separately and then concatenates them. 

Contrast to above work, another line of studies focuses more on developing unified tokenizers that fuse semantic and pixel-level features. Inspired by the success of VQGAN~\cite{esser2021taming}, early unified tokenizers predominantly adapt a discrete token design. Discrete tokenizers rely on vector quantization(VQ) to train representative unified codebooks. For example, TokenFlow~\cite{qu2025tokenflow}~jointly optimizes semantic and pixel-level features by incorporating dual codebooks with a shared mapping. VILA-U~\cite{wu2024vila} discretizes the features extracted by the SigLIP with residual quantization. UniTok~\cite{ma2025unitok}~uses multi-codebook quantization to construct unified discrete representations. 
Lately, the prevalent trend has gradually shifted toward continuous tokenizers. 
UniLIP~\cite{tang2025unilip}~tailors CLIP features for image generation by incorporating self-distillation constraints to bridge the domain gap.
Show-o2~\cite{xie2025show}~applies semantic and low-level projection to VAE latents and fuse dual features to produce unified feature space. More recently, the concurrent work, TUNA~\cite{liu2025tuna}, further simplifies this by connecting a VAE and a ViT as a unified tokenizer, which is most related to our work. However, TUNA relies on pretrained ViT checkpoints and it remains non-transparent how to train such a tokenizer. In our work, we train the ViT from scratch and propose an effective training paradigm for the unified tokenizer with unified representations.

\section{Method}
\subsection{Motivation}
Developing unified tokenizer is a pivotal step toward unifying generation and understanding, but it is often hindered by the difficulty of establishing a unified feature space and high-efficient training. Previous studies have presented impressive methods to eliminate these obstacles. However, explorations into constructing unified representations remain in their preliminary stages, and the associated training pipelines still remain non-transparent to the community. In the following, we present our model, which constructs unified vision representation space through a VAE and a ViT in a effective and straightforward way. We demonstrate to the research community how to train a unified tokenizer efficiently from scratch within the VAE latent space.

\subsection{\Model: A unified tokenizer}
\Model~uses a VAE encoder and a vision transformer(ViT) to extract unified vision features, as depicted in~\cref{fig:teaser}. The input image $x\in \mathbb{R}^{H\times W\times C}$ is first encoded by the VAE encoder $\mathcal{E}_{vae}$ from FLUX.1-dev into VAE latents $z_{vae}$, and the following training process is completely under the VAE latent space. Next, the VAE latents are fed into the ViT encoder $\mathcal{E}_{vit}$ to extract the unified representations $z_u$ for both understanding tasks and generation tasks. During the VAE stage, the FLUX.1 VAE downsamples the image height and width by 8$\times$, respectively. Therefore, we adjust the patch size of the ViT to 2$\times$2 so that the whole compression ratio is 16$\times$, which aligns with common settings. Formally,
\begin{equation}
z_{vae} = \mathcal{E}_{vae}(x) \in \mathbb{R}^{\frac{H}{8}\times \frac{W}{8}\times D_{vae}} 
\end{equation}
\begin{equation}
z_{u} = \mathcal{E}_{vit}(z_{vae}) \in \mathbb{R}^{\frac{H}{16}\times \frac{W}{16}\times D_{u}} 
\end{equation}

where $D_{vae}$ is the VAE latent channels, $D_u$ is the ViT dimensions. The encoded unified feature $z_{u}$ then goes into the reconstruction branch and the understanding branch to do decoding. \Model~employs two distinct branches to cultivate its ability to extract both generative and interpretive vision representations. The two branches are completely separate, and their respective architectures will be elaborated upon below.

\paragraph{Reconstruction branch.}
The reconstruction decoding part mirrors the structure of the tokenizer, maintaining a near-symmetrical configuration. Before the decoding, we first add noise to the unified representations in order to improve the generalization of generation ability. The perturbed feature $\tilde{z}_u$ is generated by adding Gaussian noise scaled by a sample-specific intensity:
\begin{equation}
\tilde{z}_u = z_u + \sigma \odot \epsilon, \quad \epsilon \sim \mathcal{N}(0, \mathbf{I})
\end{equation}
where $\sigma$ is uniformly sampled from $[0, \tau]$ for each instance in the batch, $\tau$ is a constant. Then we use a ViT decoder with patch size 1$\times$1 and a linear layer to convert the noised unified feature $\tilde{z}_u$ back into VAE latents $\hat{z}_{vae}$. Next, the VAE decoder is applied to decode the $\hat{z}_{vae}$ into reconstruction image $\hat{x}$. The whole reconstruction loss includes the reconstruction loss of image $\hat{x}$ and VAE latents $\hat{z}_{vae}$, and a perceptual loss based on LPIPS. The whole reconstruction loss can be formulated as:
\begin{equation}
\mathcal{L}_{rec} = {\ell}_1(x, \hat{x}) + \beta{\ell}_1(z_{vae}, \hat{z}_{vae}) + 
\lambda\mathcal{L}_{LPIPS}(x, \hat{x})
\end{equation}

\paragraph{Understanding branch.}
The paradigm of understanding branch generally follows OpenVision~\cite{li2025openvision}, where we do contrastive learning and image captioning. As shown in~\cref{fig:teaser}, we use a text encoder to extract the caption feature $z_{txt}$ to calculate contrastive loss with the unified visual feature $z_{u}$. In parallel, we utilize a text decoder to perform autoregressive prediction of synthetic captions from the unified representations and calculate the corresponding captioning loss. Formally, the understanding loss can be formulated as:
\begin{equation}
\mathcal{L}_{und} = \mathcal{L}_{caption} + \alpha\mathcal{L}_{contrastive}(z_u, z_{txt})
\end{equation}

The overall training objective is:

\begin{equation}
\mathcal{L}_{overall} = \omega_{rec}\mathcal{L}_{rec} + \omega_{und}\mathcal{L}_{und}
\end{equation}

We configure $\omega_{und}$ as double that of $\omega_{rec}$ during the training process. Reducing $\omega_{rec}$ helps to preserve generative quality while ensuring that the understanding capability remains unimpaired.

\begin{wraptable}{r}{0.5\textwidth}
    \centering
    \captionof{table}{Parameter configs for two training stages. The epoch number is defined as ImageNet-equivalent epochs (1 epoch $\approx$ 1.3M samples).}
    \resizebox{0.5\textwidth}{!}{ %
        \begin{tabular}{l|cc}
        \toprule
        Parameter & Pretraining & Finetune \\ 
        \midrule
        Resolution & 128 & 224/256 \\
        Global batch size & 8192 & 4096 \\
        Base learning rate & $8\times10^{-6}$ & $4\times10^{-7}$ \\
        Epochs & 4000 & 400 \\
        Warmup Epochs & 40 & 20 \\
        LPIPS loss weight $\lambda$ & 0 & 0.5 \\
        VAE latents $\beta$ & \multicolumn{2}{c}{0.4} \\
        Contrastive $\alpha$ & \multicolumn{2}{c}{1.0} \\
        Rec. loss $\omega_{rec}$ & \multicolumn{2}{c}{0.5} \\
        Und. loss $\omega_{und}$ & \multicolumn{2}{c}{1.0} \\
        \bottomrule
        \end{tabular}
    }
    \label{tab:parameter}
\end{wraptable}

\subsection{Training settings} 
\paragraph{Training stages and resolution.}
In accordance with the conclusions drawn in CLIPA~\cite{li2023inverse}, we employ a progressive training strategy for the tokenizer, transitioning from low-resolution to high-resolution inputs. We first pre-train the tokenizer at 128$\times$128, and then finetune it with 224$\times$224 or 256$\times$256. The epoch distribution for the two training stages is maintained at around a 10:1 ratio. By focusing most of the computation on low-resolution stages, our tokenizer attains superior performance while significantly reducing the computational overhead typically associated with high-resolution training. 

\paragraph{Training details.}
As depicted in~\cref{fig:teaser}, we use pre-trained FLUX.1 VAE and freeze it during the whole training process. All other components (including ViT encoder, ViT decoder, text encoder, text decoder, and linear layer) are randomly initialized and remain unfrozen throughout the training. For the two training stages, the global batch sizes are 8K and 4K, with cosine-decayed base learning rates of $8\times10^{-6}$ and $4\times10^{-7}$. We disable LPIPS loss during the pretraining stage to prevent loss conflicts due to varying resolutions. For complete parameter and configuration details, please refer to~\cref{tab:parameter}. The model is trained on the DataComp dataset recaptioned by LLaVA-Llama-3~\cite{li2024if}, which ensures the high quality of the training data and highly efficient multimodal learning.

\begin{table*}[t]
    \centering
    \setlength{\tabcolsep}{3pt}
    \caption{
    \textbf{Reconstruction performance of visual tokenizers.}
    Evaluations are performed on the ImageNet and COCO validation sets. Images are resized and center-cropped to 256$\times$256. Metrics includes Peak signal-to-noise ratio (PSNR), Structural Similarity Index Measure(SSIM), Learned Perceptual Image Patch Similarity (LPIPS) and reconstruction Fr\'echet inception distance (rFID).}
    \begin{tabular}{l|cccc|cccc}
    \toprule
        \multirow{2}{*}{Model} &\multicolumn{4}{c|}{ImageNet}&\multicolumn{4}{c}{COCO} \\
        \cmidrule(lr){2-5}\cmidrule(lr){6-9}
        & PSNR$\uparrow$ &SSIM$\uparrow$ &LPIPs$\downarrow$ & rFID$\downarrow$& PSNR$\uparrow$ &SSIM$\uparrow$ &LPIPs$\downarrow$ & rFID$\downarrow$  \\
    \midrule
    \multicolumn{3}{l}{\color{gray!60} \emph{Generation-oriented Tokenizer}} \\
    \color{gray!60}  SD-VAE &\color{gray!60}  26.26&\color{gray!60}  0.745 &\color{gray!60}  0.133 &\color{gray!60} 0.606&\color{gray!60}  25.99&\color{gray!60}  0.759&\color{gray!60}  0.130&\color{gray!60}  4.142 \\
\color{gray!60} SD3-VAE    & \color{gray!60} 31.29 & \color{gray!60} 0.886 & \color{gray!60} 0.059 & \color{gray!60} 0.201 & \color{gray!60} 31.18& \color{gray!60} 0.894& \color{gray!60} 0.056& \color{gray!60} 1.671  \\
\color{gray!60}Cosmos &\color{gray!60} 25.07&\color{gray!60} 0.700&\color{gray!60} 0.167&\color{gray!60} 0.959&\color{gray!60} 24.74&\color{gray!60} 0.711&\color{gray!60} 0.165&\color{gray!60} 5.063\\
        \color{gray!60} FLUX-VAE   & \color{gray!60} 32.86 & \color{gray!60} 0.917 & \color{gray!60} 0.044 & \color{gray!60} 0.176 & \color{gray!60} 32.73 & \color{gray!60}0.923& \color{gray!60} 0.041& \color{gray!60} 1.343 \\
        \color{gray!60} Wan2.1-VAE & \color{gray!60} 31.34 & \color{gray!60} 0.886 & \color{gray!60} 0.058 & \color{gray!60} 0.945  & \color{gray!60}31.19 & \color{gray!60} 0.895 & \color{gray!60} 0.055 & \color{gray!60} 3.449 \\
        
     \midrule
    \multicolumn{3}{l}{\emph{Unified Tokenizer}} \\
    RAE (CLIP)& 17.44&0.403&0.324&1.06 &16.98&0.394&0.345&10.119 \\
        UniTok &25.34& 0.742& 0.132& 0.362 & 24.95 & 0.750 & 0.131 & 3.918\\
        OmniTokenizer&24.69& 0.771& 0.138& 1.411&24.31& 0.779& 0.137& 6.292 \\
        Vila-U & 22.24 &0.612& 0.228 &4.231 &  21.89& 0.620& 0.227& 10.997\\
        \rowcolor{green!15}
         \Model& 30.92&0.902&0.053&0.187&30.89&0.907&0.050&1.601 \\  
    \bottomrule
    \end{tabular}
    \label{tab:reconstruction}
\end{table*}

\section{Experiments}

\subsection{Evaluation settings}
To comprehensively evaluate the performance of our unified tokenizer, we evaluate the reconstruction, generation and understanding performance and report their results in~\cref{sec:reconstruction}. For the generation side, we follow RAE configs to train a generative model with DiT and a wide DDT head, and evaluate the generation fedelity of~\Model~on ImageNet. For the understanding side, we train vision-language models with our tokenizer under LLaVA-1.5 and LLaVA-NeXT frameworks~\cite{liu2024improved}, and evaluate the understanding performance across a range of downstream multimodal benchmarks.

\begin{figure}[t]
    \centering
    \includegraphics[width=\linewidth]{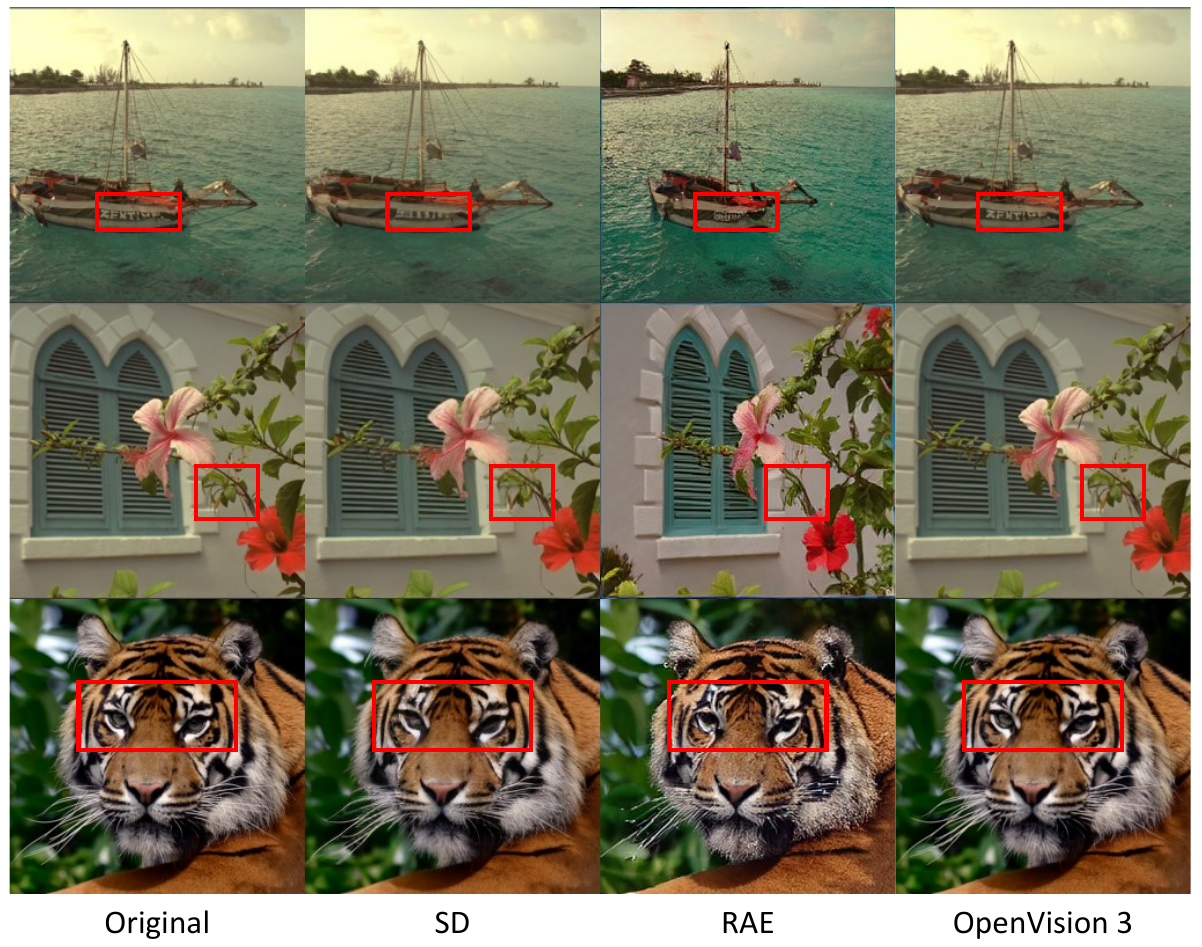}
    \caption{
    \textbf{Reconstruction visualization.}
    We present three cases of image reconstruction to compare~\Model~with SD-VAE and RAE. Our tokenizer excels at preserving textual content (Case 1) and intricate image details (Case 2 and 3). There is hardly any perceptible difference between our reconstructions and the ground truth.
    }
    \label{fig:recon_case}
\end{figure}

\subsection{Reconstruction performance}
\label{sec:reconstruction}
As shown in~\cref{tab:reconstruction}, \Model~significantly outperforms existing unified tokenizers across all metrics. Previous unified models~\cite{zheng2025diffusion, ma2025unitok, wang2024omnitokenizer, wu2024vila}~often struggle to maintain high reconstruction quality due to the trade-off required to align with semantic objectives (\eg, SigLIP alignment). For instance, on ImageNet, \Model~ achieves a PSNR of 30.92 dB, surpassing UniTok (25.34 dB) and Vila-U (22.24 dB) by a wide margin. Similarly, in terms of perceptual quality, our model achieves an LPIPS score of 0.053, whereas the closest unified competitor, UniTok, lags behind at 0.132. 
This demonstrates that our architecture—utilizing the VAE-ViT hybrid design—successfully mitigates the information loss typically associated with semantic compression. 
On the COCO dataset, our tokenizer keeps a substantial performance advantage. For instance, our rFID significantly outperforms UniTok (1.601~\vs~3.918) and other unified tokenizers by a large margin, further demonstrating our robust generalizability.
Notably, even in comparison with specialized generation-oriented tokenizers~\cite{rombach2022high, esser2024scaling, agarwal2025cosmos, labs2025flux, wan2025wan}, our model maintains competitive or better results.

\paragraph{Visualization.}
To provide a visual demonstration of our reconstruction performance, we present several qualitative results in~\cref{fig:recon_case}. Compared to SD-VAE and RAE,~\Model~exhibits superiority in preserving intricate image details and faithfully reconstructing text. 
In the first case, our tokenizer fully recovers the characters on the hull, whereas both SD-VAE and RAE fail. 
In the second and third cases, our tokenizer exhibits superiority in capturing fine-grained details, such as the textures of flowers and foliage and the intricacies of the tiger’s eyes, demonstrating a higher-fidelity reconstruction capability.

\begin{figure}[t]
    \centering
    
    \begin{minipage}{0.56\textwidth}
        \centering
        \captionof{table}{
        \textbf{Class-conditional image generation on ImageNet 256x256.} We report gFID, Inception Score (IS), Precision (Pre.), and Recall (Rec.). Our tokenizer achieves the best performance across all metrics.}
        \resizebox{\linewidth}{!}{
            \begin{tabular}{l|c|cccc}
                \toprule
                Tokenizer & Generator & gFID$\downarrow$  & IS$\uparrow$ & Pre.$\uparrow$ & Rec.$\uparrow$ \\
                \midrule
                SD-VAE & DiT & 2.27  & 278.2 & 0.83 & 0.57 \\
                SD-VAE & SiT & 2.06 & 270.3 & 0.82 & \textbf{0.59} \\
                UniTok & LlamaGen & 2.51 & 216.7 & 0.82 & 0.57 \\
                CLIP & RAE & 2.54 & 256.4 & 0.80 & 0.54 \\
                OpenVision & RAE & 2.44 & 262.2 & 0.80 & 0.53 \\
                \rowcolor{green!15} 
                \Model & RAE & \textbf{1.87} & \textbf{290.0} & \textbf{0.84} & \textbf{0.59} \\
                \bottomrule
            \end{tabular}
        }
        \label{tab:generation}
    \end{minipage}
    \hfill %
    \begin{minipage}{0.4\textwidth}
        \centering
        \includegraphics[width=\linewidth]{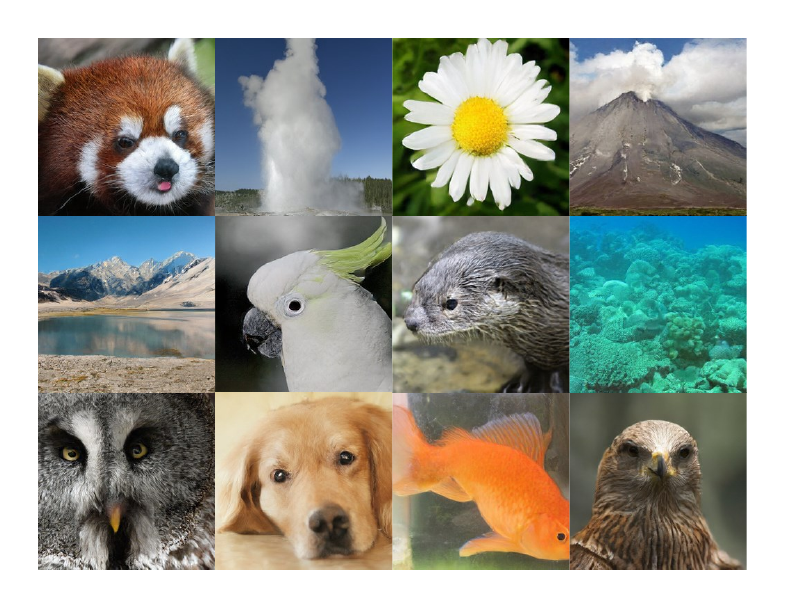}
        \caption{\textbf{Qualitative results of class-conditional ImageNet-256 generation.} Under the RAE framework,~\Model~is able to generate high quality images.}
        \label{fig:gen}
    \end{minipage}
\end{figure}

\subsection{Generation performance}
\label{sec:generation}
As shown in we report generation Fr\'echet inception distance (gFID), Inception Score (IS), Precision (Pre.), and Recall (Rec.) as evaluation metric. 
We present the generative performance of each tokenizer when paired with its respective compatible generator. For low-level tokenizer, we evaluate SD-VAE with traditional diffusion-based generative models (DiT and SiT)~\cite{peebles2023scalable, ma2024sit}. For semantic tokenizer, we select CLIP with RAE generator for fair comparison with our tokenizer. According to~\cref{tab:generation},~\Model~outperforms these tokenizers across all the metrics. For example, we achieve better gFID when compared to SD-VAE with improved generator SiT(1.87~\vs~2.06). Our tokenizer also surpasses semantic encoders like CLIP~\cite{clip}~by a large margin in generation(1.87~\vs~2.54).

\paragraph{Visualization.} As shown in~\cref{fig:gen}, we visualize some class-conditional ImageNet-256 generation results with~\Model~under RAE framework. The images generated by our tokenizer exhibit structurally coherent objects with rich stylistic details, which shows the strong capability of our tokenizer in generating high-quality samples with great fidelity and diversity. 

\subsection{Understanding performance}
\label{sec:understanding}
To evaluate the semantic representation capability of~\Model, we integrate it into the LLaVA-1.5 and LLaVA-NeXT framework and conduct training following their standard training configurations. Due to the fixed downsample size of VAE, we keep the same encoded token numbers with OpenAI CLIP for fair comparison.
In~\cref{Tab:understanding}~and~\cref{Tab:und_next}, we compare our tokenizer with  CLIP and present the results on multiple multimodal benchmarks, including MME~\cite{mme}, ScienceQA~\cite{scienceqa}, SeedBench~\cite{seedbench}, GQA~\cite{gqa}~and POPE~\cite{pope}.
According to the tables,~\Model~can match or exceed the understanding performance of CLIP on different model sizes (Base and Large).
For example, our tokenizer consistently surpasses CLIP on SeedBench (63.1~\vs~62.2 and 65.8~\vs~65.4) and POPE (83.7~\vs~82.9 and 85.2~\vs~84.7) under LLaVA-1.5 framework. 
In the stronger LLaVA-NeXT setting,~\Model~maintains its performance advantage over CLIP.
For instance, we present pronounced performance gap over the CLIP baseline on SeedBench (63.3~\vs~61.2 and 68.6~\vs~61.8) and GQA (59.2~\vs~58.1 and 62.0~\vs~59.4), exhibiting its exceptional robustness.

The results strongly demonstrate that our unified tokenizer is comparable to the understanding-oriented tokenizer CLIP in terms of semantic comprehension, and even displays a clear advantage in certain aspects.

\begin{table*}[t!]
    \centering
    \caption{
    \textbf{Comparison of~\Model~with OpenAI CLIP under LLaVA-1.5 framework.} We evaluate the understanding performance of our tokenizer and CLIP on multiple multimodal benchmarks. With the same image token numbers, our unified tokenizer performs on par with OpenAI CLIP under both Base and Large size, while surpassing CLIP across some specific benchmarks. 
    }
    \label{Tab:understanding}
    \resizebox{\linewidth}{!}{
    \begin{tabular}{c|c|c|c|c|c|c|c|c|c}
    \toprule
    Method &
    Vision Encoder &
    \# Tokens &
    \# Res. &
    MME-P &
    MME-C &
    SeedBench &
    ScienceQA &
    GQA  &
    POPE \\
    \midrule
    OpenAI-CLIP & B/16 & 196 & 224 & \textbf{1399} & \textbf{318} & 62.2 & \textbf{73.7} & 58.6 & 82.9 \\
    \rowcolor{green!15} 
    \Model & VAE + B/2 & 196 & 224 & 1388 & 275 & \textbf{63.1} & 73.1 & \textbf{58.9} & \textbf{83.7} \\

    \midrule
    
    OpenAI-CLIP & L/14 & 256 & 224 &\textbf{1468} & 292 & 65.4 & \textbf{73.9}  & 60.6& 84.7 \\    
    
    \rowcolor{green!15} 
    \Model & VAE + L/2 & 256 & 256  & 1405 & \textbf{313} & \textbf{65.8} & 72.4 & \textbf{61.0} & \textbf{85.2}  \\

    \bottomrule
    \end{tabular}}
\end{table*}

\begin{table*}[t!]
    \centering
    \caption{
    \textbf{Comparison of~\Model~with OpenAI CLIP under LLaVA-NeXT framework.} We train~\Model~and OpenAI CLIP under LLaVA-NeXT setting. The results reveal that our tokenizer show competitive understanding performance with CLIP. 
    }
    \label{Tab:und_next}
    \resizebox{\linewidth}{!}{
    \begin{tabular}{c|c|c|c|c|c|c|c|c|c}
    \toprule
    Method &
    Vision Encoder &
    \# Tokens &
    \# Res. &
    MME-P &
    MME-C &
    SeedBench &
    ScienceQA &
    GQA  &
    POPE \\
    \midrule
    OpenAI-CLIP & B/16 & 196 & 224 & \textbf{1368} & 316 & 61.2 & \textbf{72.8} & 58.1 & 84.7 \\
    \rowcolor{green!15}
    \Model & VAE + B/2 & 196 & 224 & 1340 & \textbf{337} & \textbf{63.3} & 68.9 & \textbf{59.2} & \textbf{84.9} \\

    \midrule
    
    OpenAI-CLIP & L/14 & 256 & 224 & 1446 & \textbf{319} & 61.8 & \textbf{75.3} & 59.4 & 84.1 \\    
    
    \rowcolor{green!15} 
    \Model & VAE + L/2 & 256 & 256 & \textbf{1449} & 280 & \textbf{68.6} & 73.6 & \textbf{62.0} & \textbf{86.6} \\

    \bottomrule
    \end{tabular}}
\end{table*}

\begin{figure*}[t]
    \centering
    \subfloat[pixel recon loss]{
      \begin{minipage}[b]{0.23\textwidth}
        \includegraphics[width=1\textwidth]{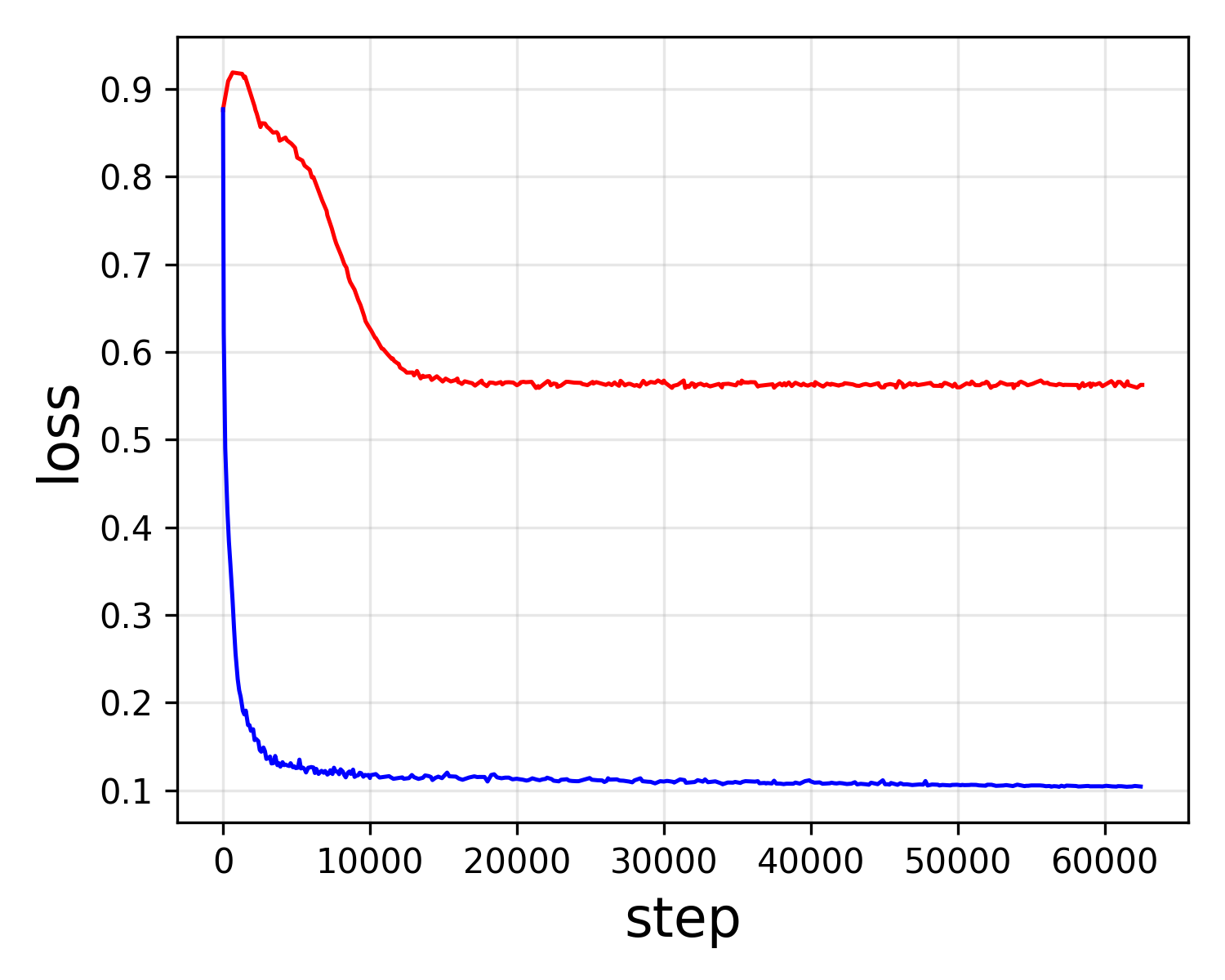}
      \end{minipage}
      \label{fig:pixel_loss}
      }
    \subfloat[latents recon loss]{
      \begin{minipage}[b]{0.23\textwidth}
        \includegraphics[width=1\textwidth]{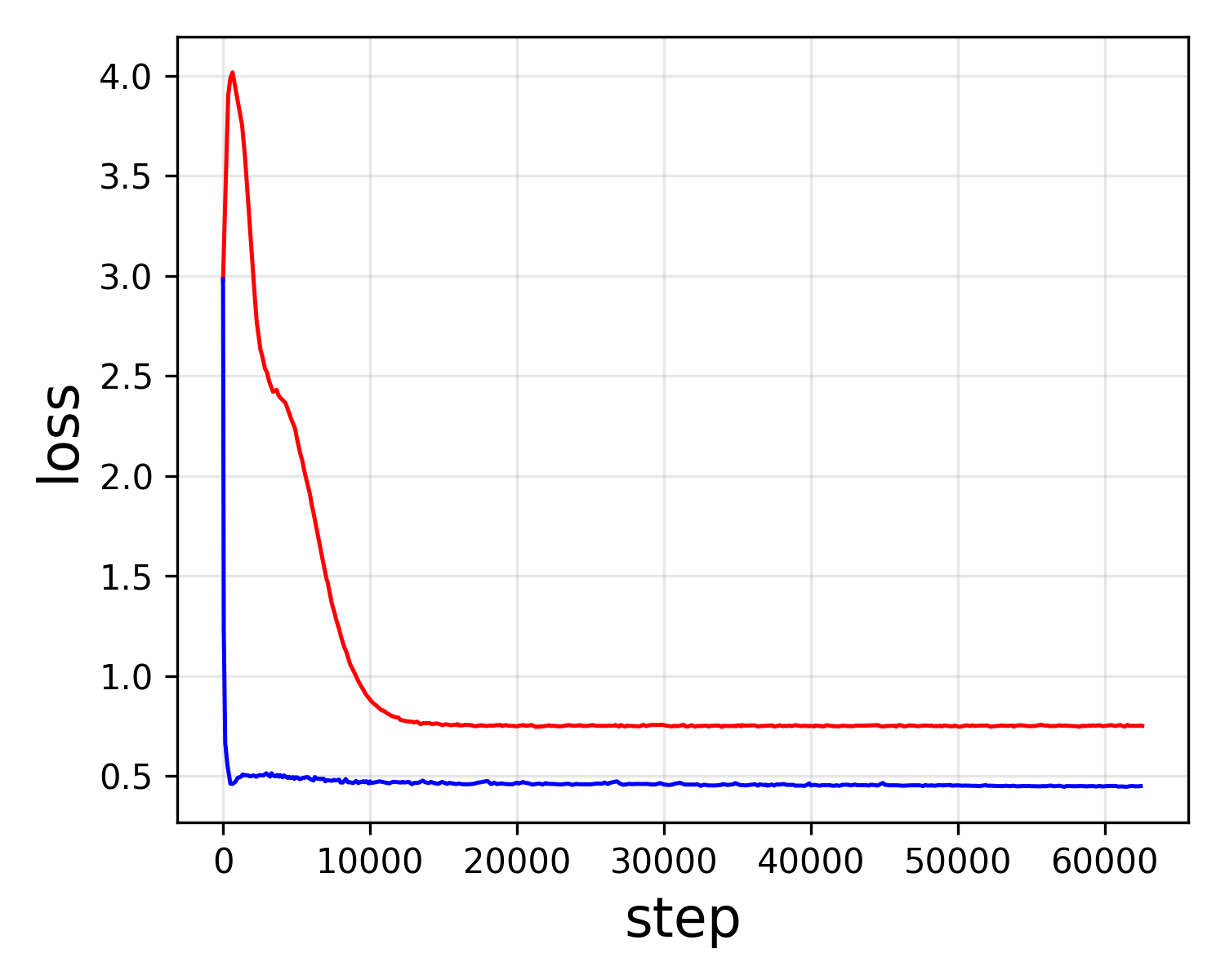}
      \end{minipage}
      \label{fig:latent_loss}
      }
    \subfloat[caption loss]{
      \begin{minipage}[b]{0.23\textwidth}
        \includegraphics[width=1\textwidth]{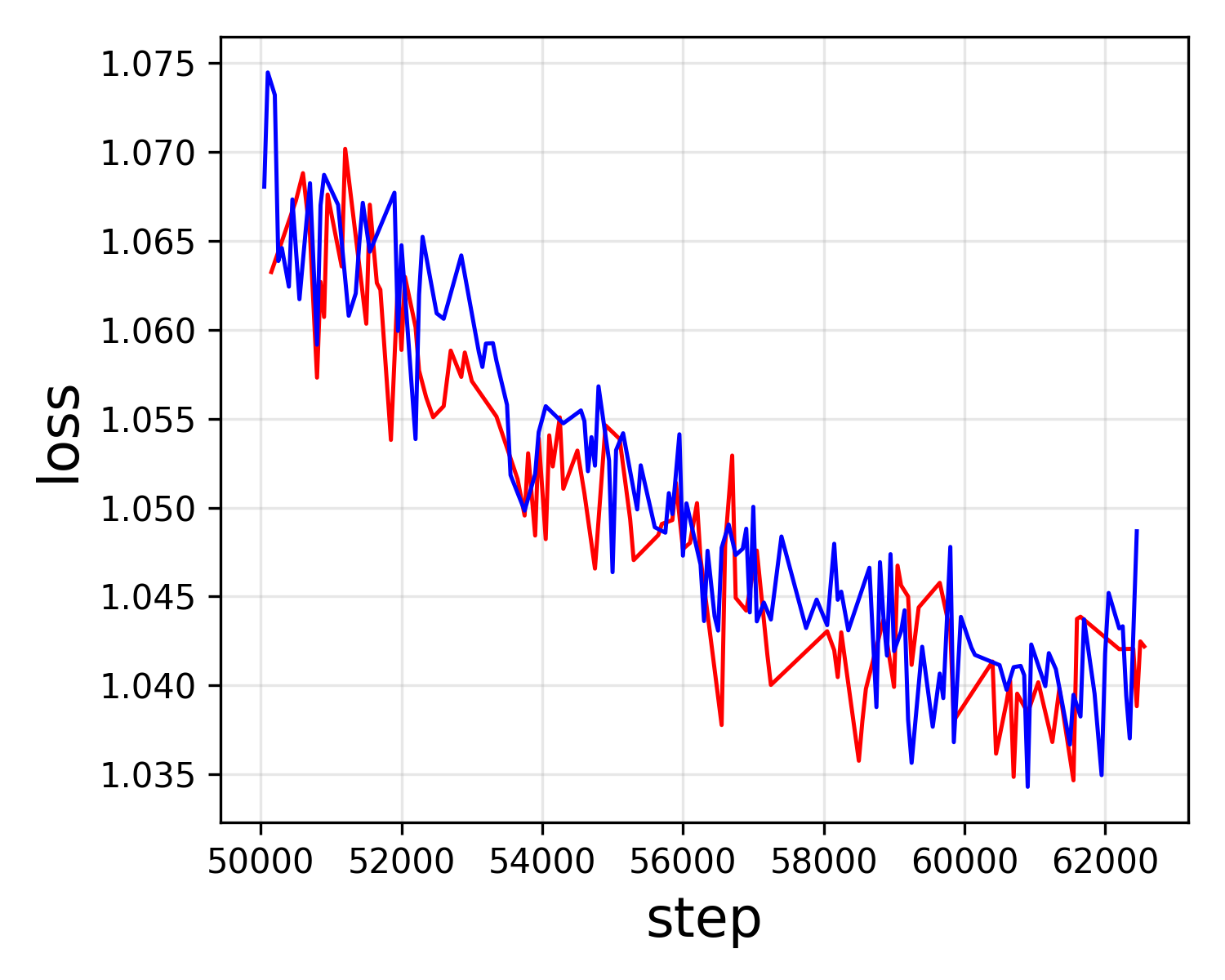}
      \end{minipage}
  \label{fig:caption_loss}
      }
      \subfloat[contrastive loss]{
      \begin{minipage}[b]{0.23\textwidth}
        \includegraphics[width=1\textwidth]{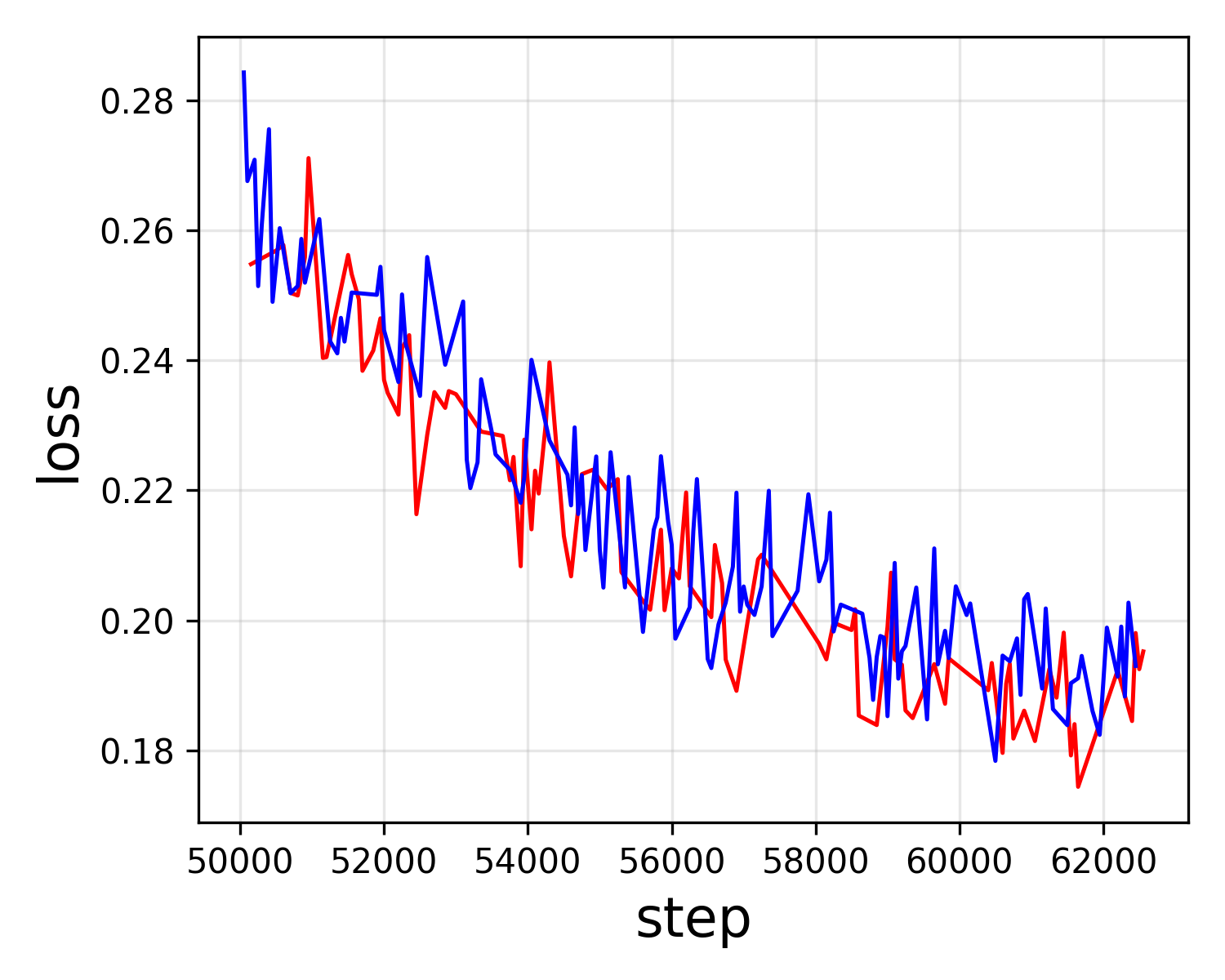}
      \end{minipage}
      \label{fig:contrastive_loss}
      }
    \caption{
          \textbf{Loss visualization with only semantic loss.} 
    We trained our tokenizer \textcolor{blue}{with} and \textcolor{red}{without} the reconstruction loss, respectively. 
    In Figures (a) and (b), both pixel-level and latent-level reconstruction losses decrease significantly even in the absence of explicit reconstruction signals. Figures (c) and (d) demonstrate that the incorporation of the reconstruction loss has no adverse impact on the losses of the understanding branch.
          }
    \label{fig:only_und}
\end{figure*}

\begin{figure*}[t]
    \centering
    \subfloat[pixel recon loss]{
      \begin{minipage}[b]{0.23\textwidth}
        \includegraphics[width=1\textwidth]{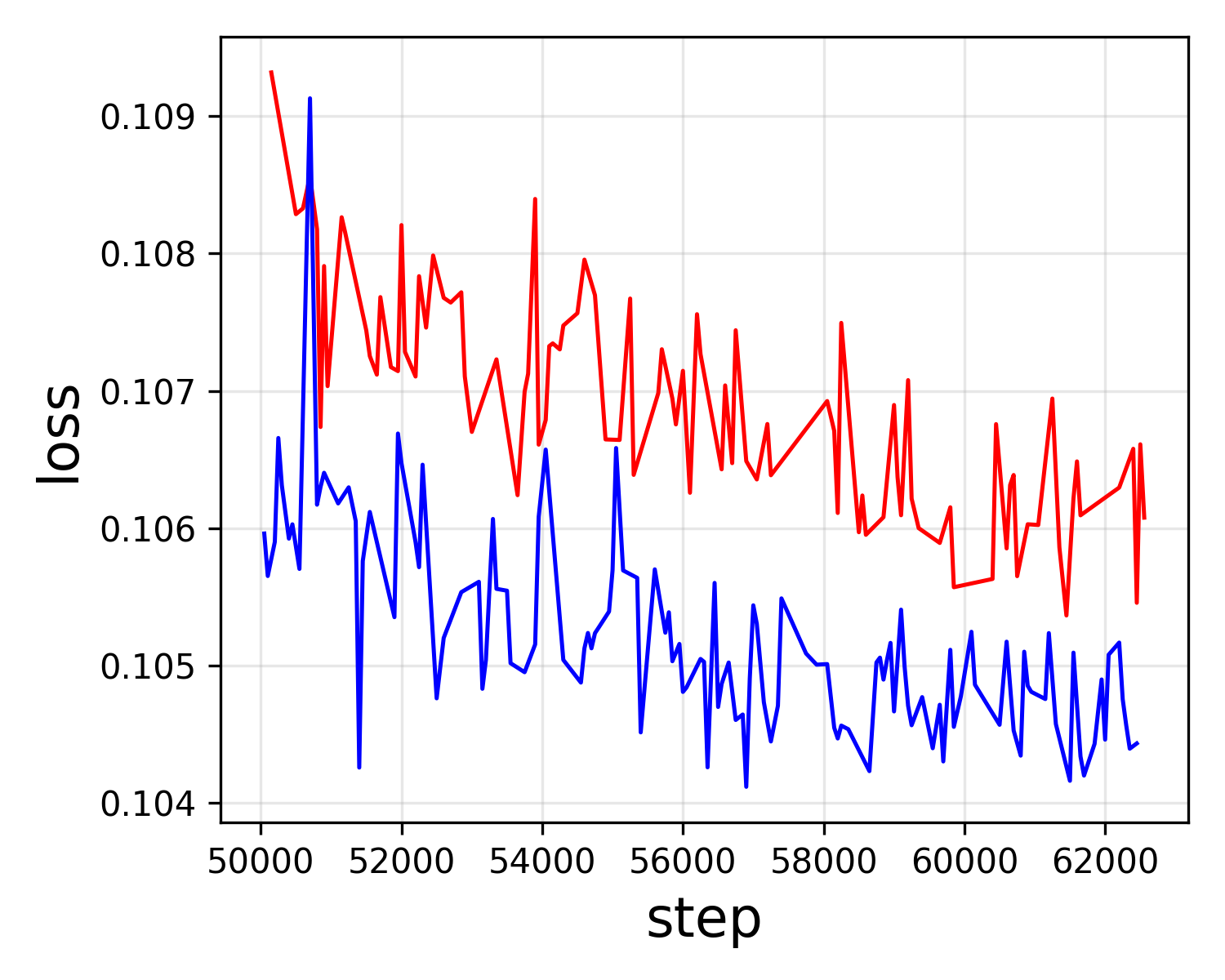}
      \end{minipage}
      \label{fig:only_rec_pixel_loss}
      }
    \subfloat[latents recon loss]{
      \begin{minipage}[b]{0.23\textwidth}
        \includegraphics[width=1\textwidth]{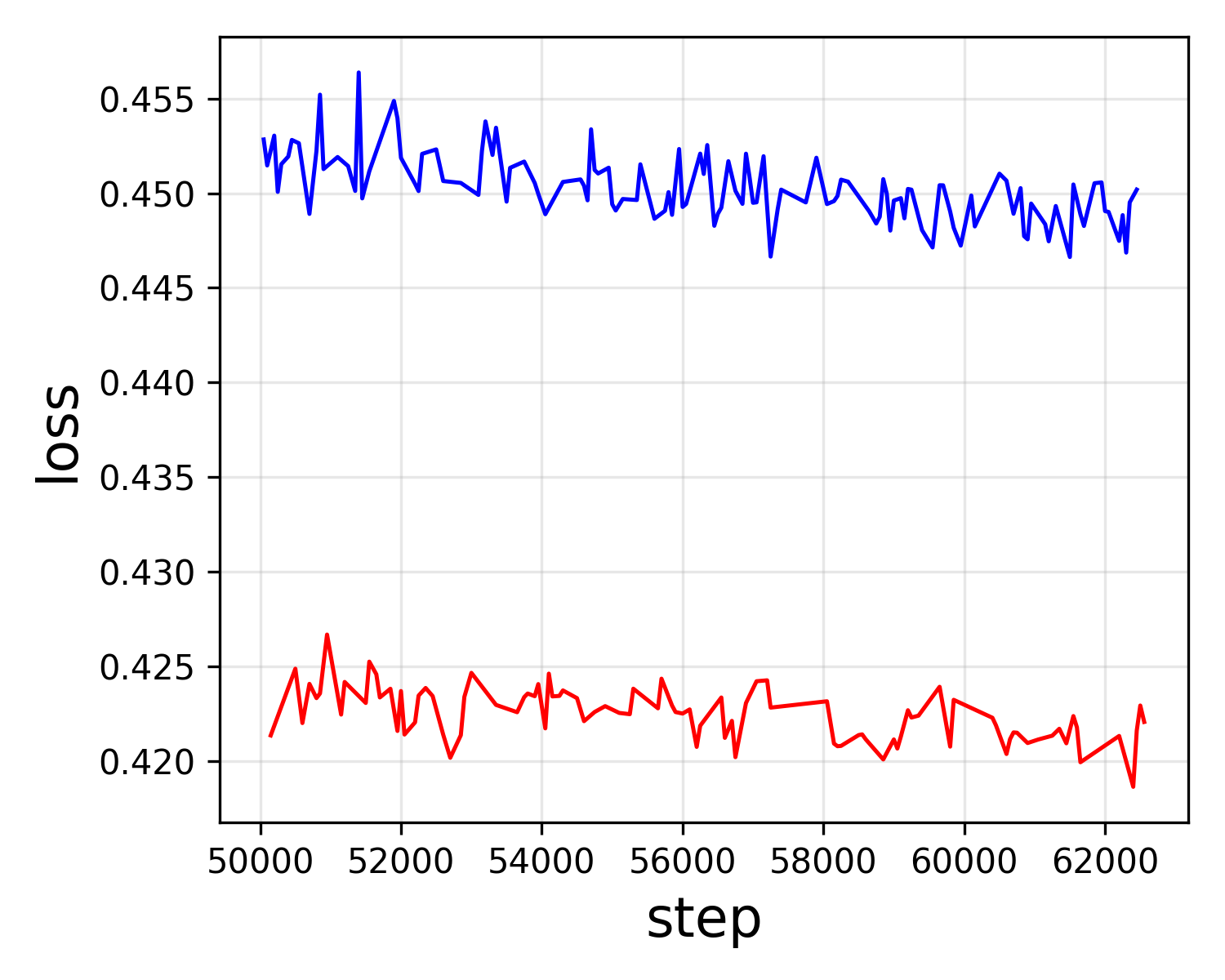}
      \end{minipage}
      \label{fig:only_rec_latent_loss}
      }
    \subfloat[caption loss]{
      \begin{minipage}[b]{0.23\textwidth}
        \includegraphics[width=1\textwidth]{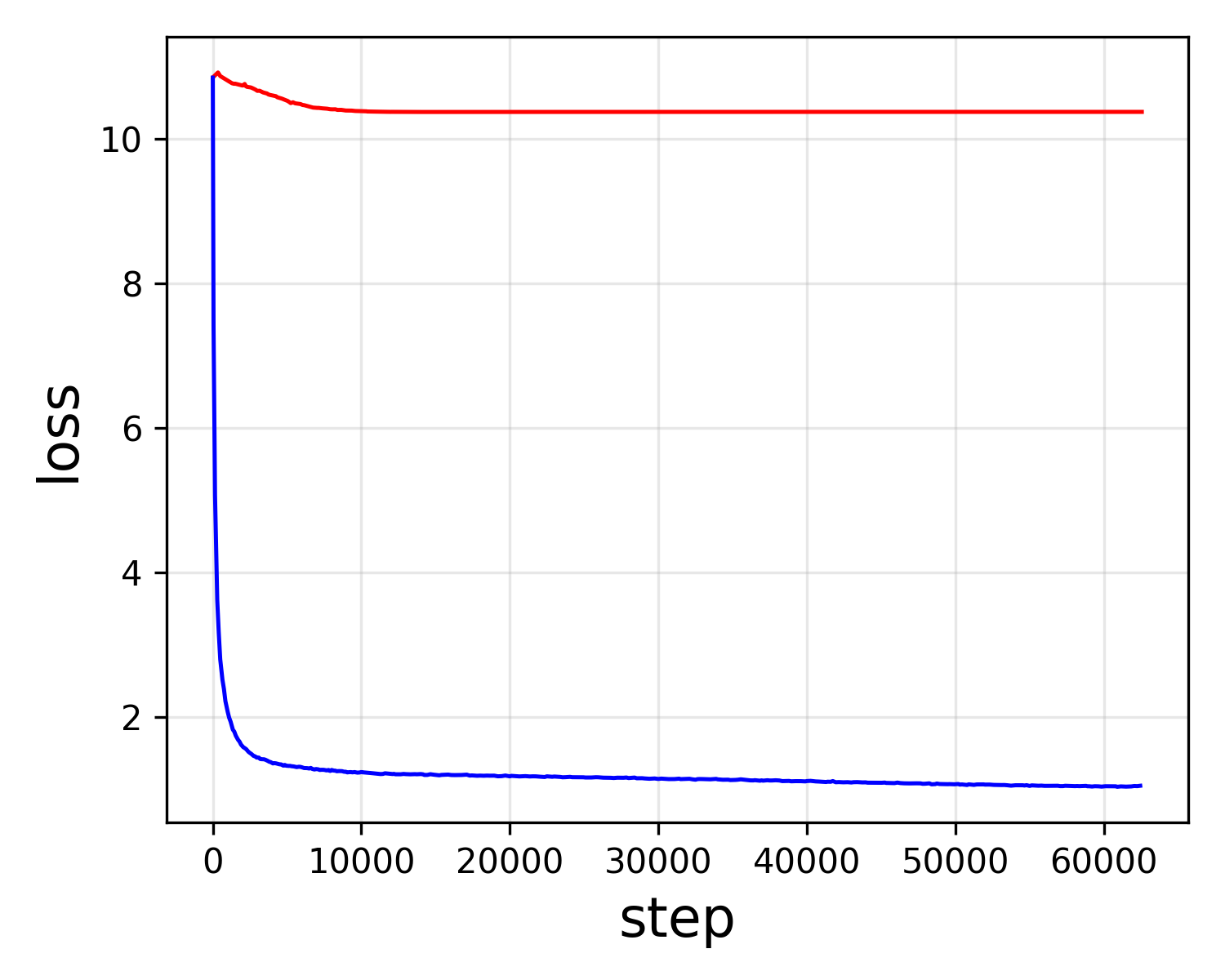}
      \end{minipage}
  \label{fig:only_rec_caption_loss}
      }
      \subfloat[contrastive loss]{
      \begin{minipage}[b]{0.23\textwidth}
        \includegraphics[width=1\textwidth]{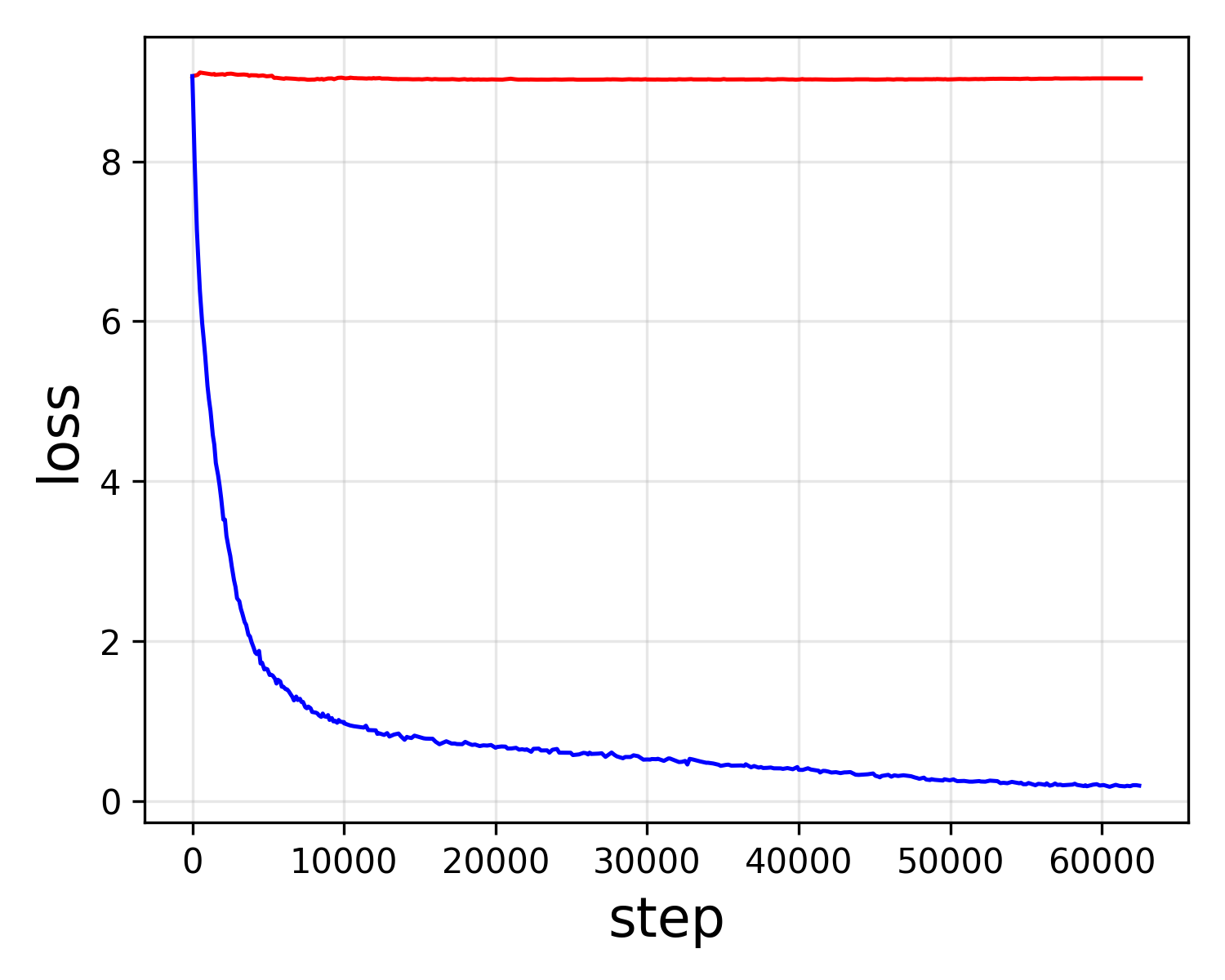}
      \end{minipage}
      \label{fig:only_rec_contrastive_loss}
      }
    \caption{
    \textbf{Loss visualization with only reconstruction loss.} 
    We trained our tokenizer \textcolor{blue}{with} and \textcolor{red}{without} the understanding loss, respectively. 
    In Figure (a), the inclusion of semantic loss leads to a lower image reconstruction loss, suggesting that semantic supervision can, in turn, enhance reconstruction performance. Figures (c) and (d) reveal that the caption loss exhibits a slight downward trend despite the lack of direct semantic signals, and the contrastive loss remains almost stagnant.
        }
    \label{fig:only_rec}
\end{figure*}

\section{Discussion}

\subsection{Reciprocal synergy between understanding and reconstruction}
\label{sec:loss_interaction}
For unified tokenizers, balancing the capabilities of understanding and generation remains a long-standing challenge. 
To investigate the mutual influence of these two objectives within our tokenizer, we conduct experiments by training the model exclusively with the understanding loss and exclusively with the reconstruction loss, respectively.

\paragraph{Remove reconstruction loss.}
In~\cref{fig:only_und}, we remove reconstruction loss and train only with semantic loss. The blue curve represents the baseline loss, while the red curve denotes the model trained without the reconstruction loss. According to the loss curves in~\cref{fig:pixel_loss}~and~\cref{fig:latent_loss}, even in the absence of reconstruction objectives, the reconstruction loss still exhibits a substantial decline, suggesting that our semantic objectives contribute significantly to image reconstruction. Furthermore, comparing the red and blue curves in~\cref{fig:caption_loss}~and~\cref{fig:contrastive_loss}, it is evident that the incorporation of the reconstruction loss leads to no significant change in either caption or contrastive loss. These observations collectively indicate a mutually beneficial synergy between the two types of losses.

\paragraph{Remove understanding loss.}
In~\cref{fig:only_rec}, we remove understanding loss and train only with reconstruction-driven signals. The red curves here denote the loss without reconstruction-driven signals. 
~\cref{fig:only_rec_caption_loss}~and~\cref{fig:only_rec_contrastive_loss}~show that in the absence of semantic supervision, the contrastive loss remains almost stagnant, whereas the caption loss exhibits a marginal decline.
This indicates that the reconstruction task intrinsically facilitates semantic tasks that are also generative in nature.
Moreover, as seen in~\cref{fig:only_rec_pixel_loss}, the addition of semantic loss paradoxically improves reconstruction performance, providing further evidence of the synergistic relationship between these two branches.

\begin{table}[t]
    \centering
    \small
    \begin{minipage}[t]{0.64\textwidth}
        \centering
        \caption{
        \textbf{Effectiveness of VAE in reconstruction.} 
        We study the influence of VAE latents in reconstruction performance. 
        Compared to the variant without a VAE,~\Model~demonstrates a significant advantage in rFID, achieving superior reconstruction quality. Our tokenizer maintains competitive results on PSNR, SSIM, and LPIPS with slight differences.
        }
        \setlength{\tabcolsep}{4pt}
        \begin{tabular}{l|cccc}
        \toprule
         Model   & PSNR$\uparrow$ &SSIM$\uparrow$ &LPIPs$\downarrow$ & rFID$\downarrow$  \\
            \midrule
        
       w/o VAE & 32.82 & 0.935 & 0.060&0.980 \\
        \rowcolor{green!15}
        \Model &30.33&0.885  &0.061 &0.216\\
        \bottomrule
        \end{tabular}
        \label{tab:recon_vae}
    \end{minipage}
    \hfill
    \begin{minipage}[t]{0.32\textwidth}
        \centering
        \caption{
        \textbf{Effectiveness of VAE in Generation.} 
        Under the VAE latent space,~\Model~outperforms its counterpart by 1.23 gFID, showing better generation ability.
        }
        \setlength{\tabcolsep}{6pt} %
        \begin{tabular}{l|c}
        \toprule
         Model   & gFID$\downarrow$  \\
        \midrule
        w/o VAE & 9.68 \\
        \rowcolor{green!15}
        \Model &8.45\\
        \bottomrule
        \end{tabular}
        \label{tab:gen_vae}
    \end{minipage}
\end{table}

\begin{table*}[t!]
    \centering
    \caption{
    \textbf{Understanding performance comparison of unified tokenizer with and without VAE.}
    We evaluate the multimodal understanding capabilities of~\Model~and its non-VAE counterpart under the LLaVA-1.5 and LLaVA-NeXT architectures. Under identical training settings, it can be observed that~\Model~consistently outperforms the non-VAE version. Under the LLaVA-1.5 framework, our tokenizer outperforms the counterpart in four out of six metrics. Notably, under the LLaVA-NeXT framework, our model achieves a comprehensive lead across all metrics.
    }
    \resizebox{\linewidth}{!}{
    \begin{tabular}{c|c|c|c|c|c|c|c|c|c}
    \toprule
    Method &
    Vision Encoder &
    \# Tokens &
    \# Res. &
    MME-P &
    MME-C &
    SeedBench &
    ScienceQA &
    GQA  &
    POPE \\
    \midrule
     \multicolumn{10}{l}{\color{gray!60} \emph{LLaVA-1.5 Framework}} \\
    \rowcolor{gray!15}
    \Model & VAE + B/2 & 196 & 224 & \textbf{1382} & 287 & \textbf{62.4} & \textbf{73.0} & \textbf{58.0} & 83.7 \\
    w/o VAE & B/16 & 196 & 224 & 1366 & \textbf{307} & 62.2 & 72.7 & 57.6 & \textbf{83.8} \\
    
    \midrule
     \multicolumn{10}{l}{\color{gray!60} \emph{LLaVA-NeXT Framework}} \\
    \rowcolor{gray!15}
    \Model & VAE + B/2 & 196 & 224 & \textbf{1383} & \textbf{348} & \textbf{63.6} & \textbf{70.5} & \textbf{59.1} & \textbf{84.4} \\
    w/o VAE & B/16 & 196 & 224 & 1346 & 296 & 62.0 & 67.8 & 57.7 & 83.8 \\

    \bottomrule
    \end{tabular}}
    \label{Tab:wo_vae_und}
\end{table*}

\begin{figure*}[t]
    \centering
    \subfloat[pixel recon loss]{
      \begin{minipage}[b]{0.23\textwidth}
        \includegraphics[width=1\textwidth]{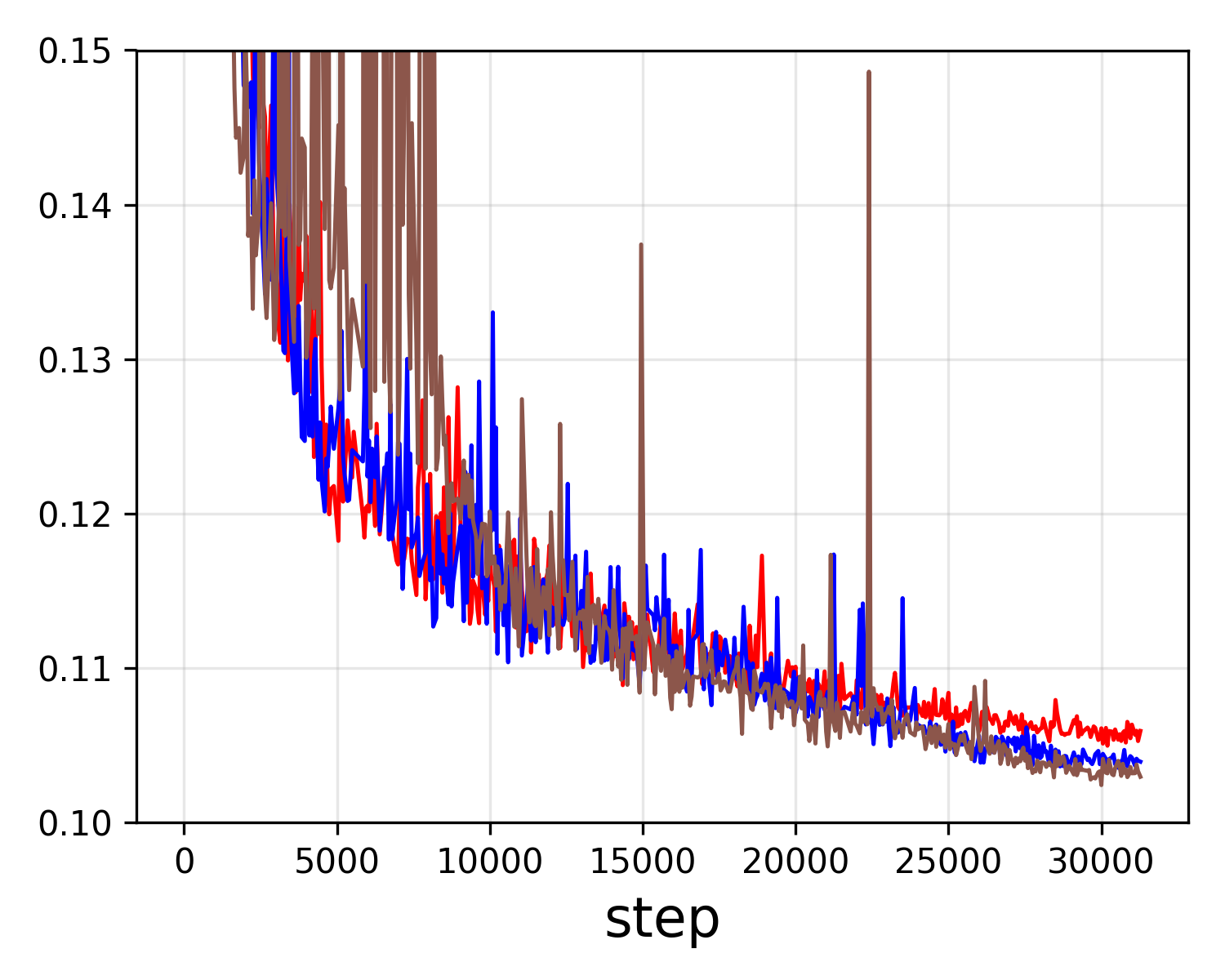}
      \end{minipage}
      \label{fig:dec_pixel_loss}
      }
    \subfloat[latents recon loss]{
      \begin{minipage}[b]{0.23\textwidth}
        \includegraphics[width=1\textwidth]{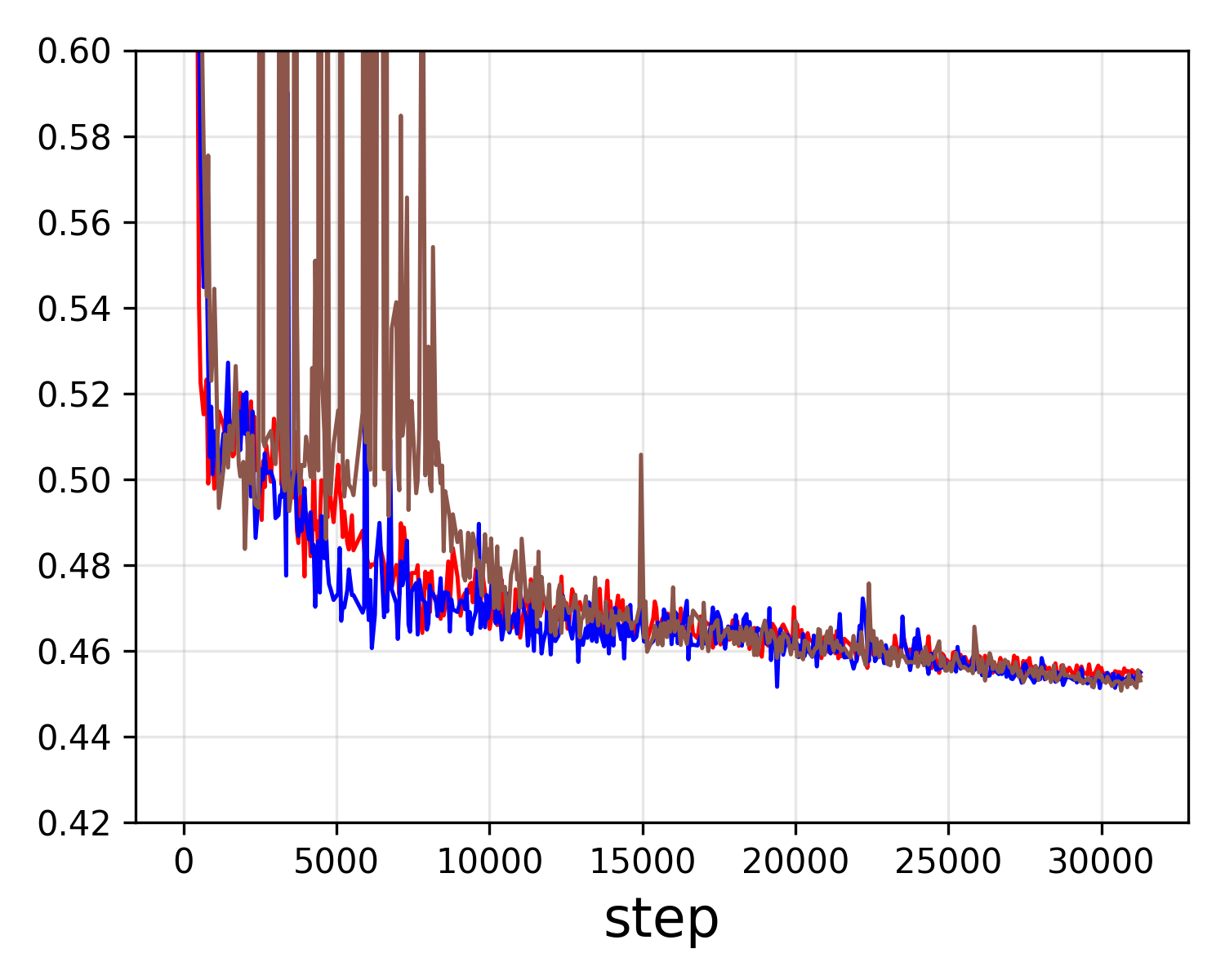}
      \end{minipage}
      \label{fig:dec_latent_loss}
      }
    \subfloat[caption loss]{
      \begin{minipage}[b]{0.23\textwidth}
        \includegraphics[width=1\textwidth]{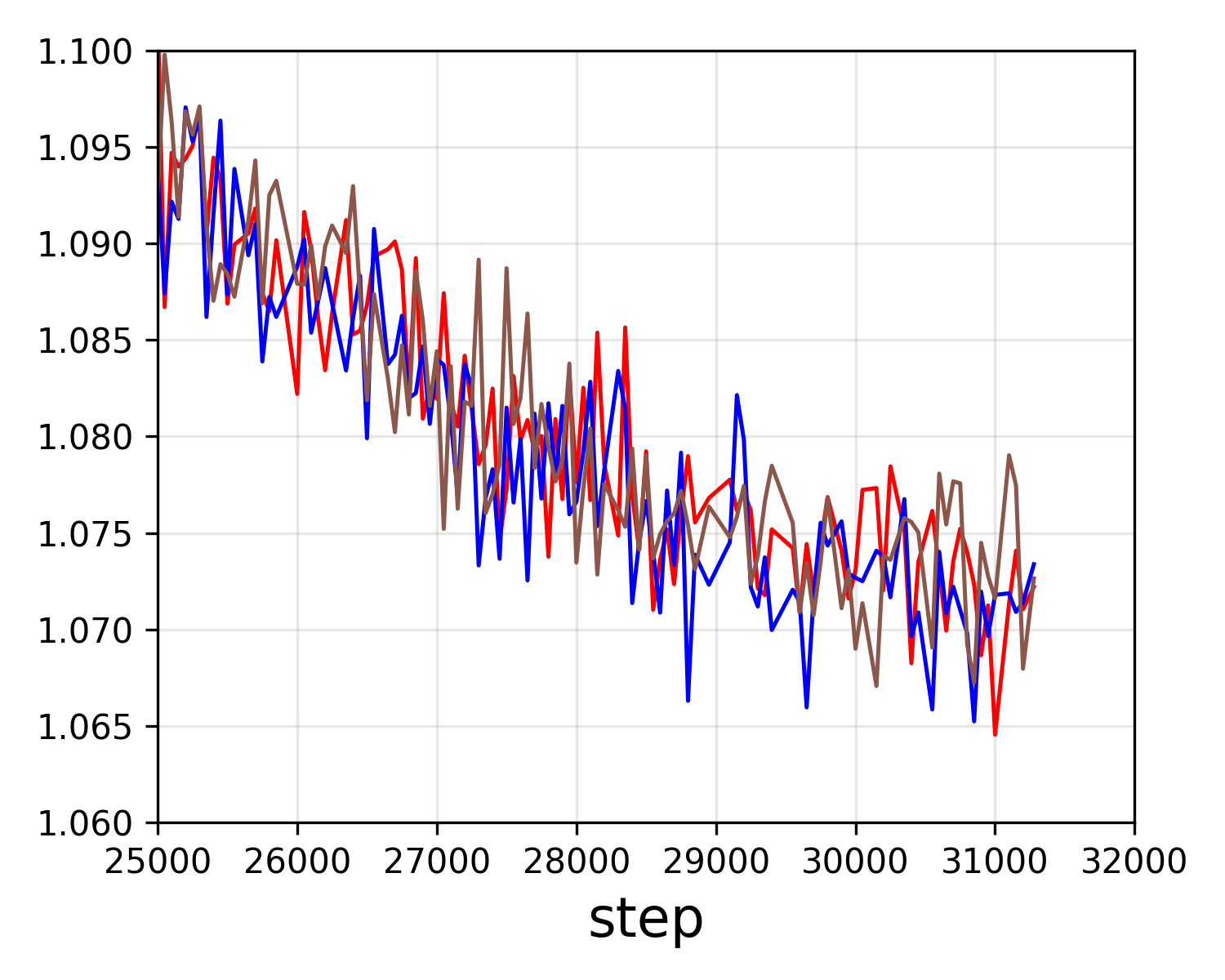}
      \end{minipage}
  \label{fig:dec_caption_loss}
      }
      \subfloat[contrastive loss]{
      \begin{minipage}[b]{0.23\textwidth}
        \includegraphics[width=1\textwidth]{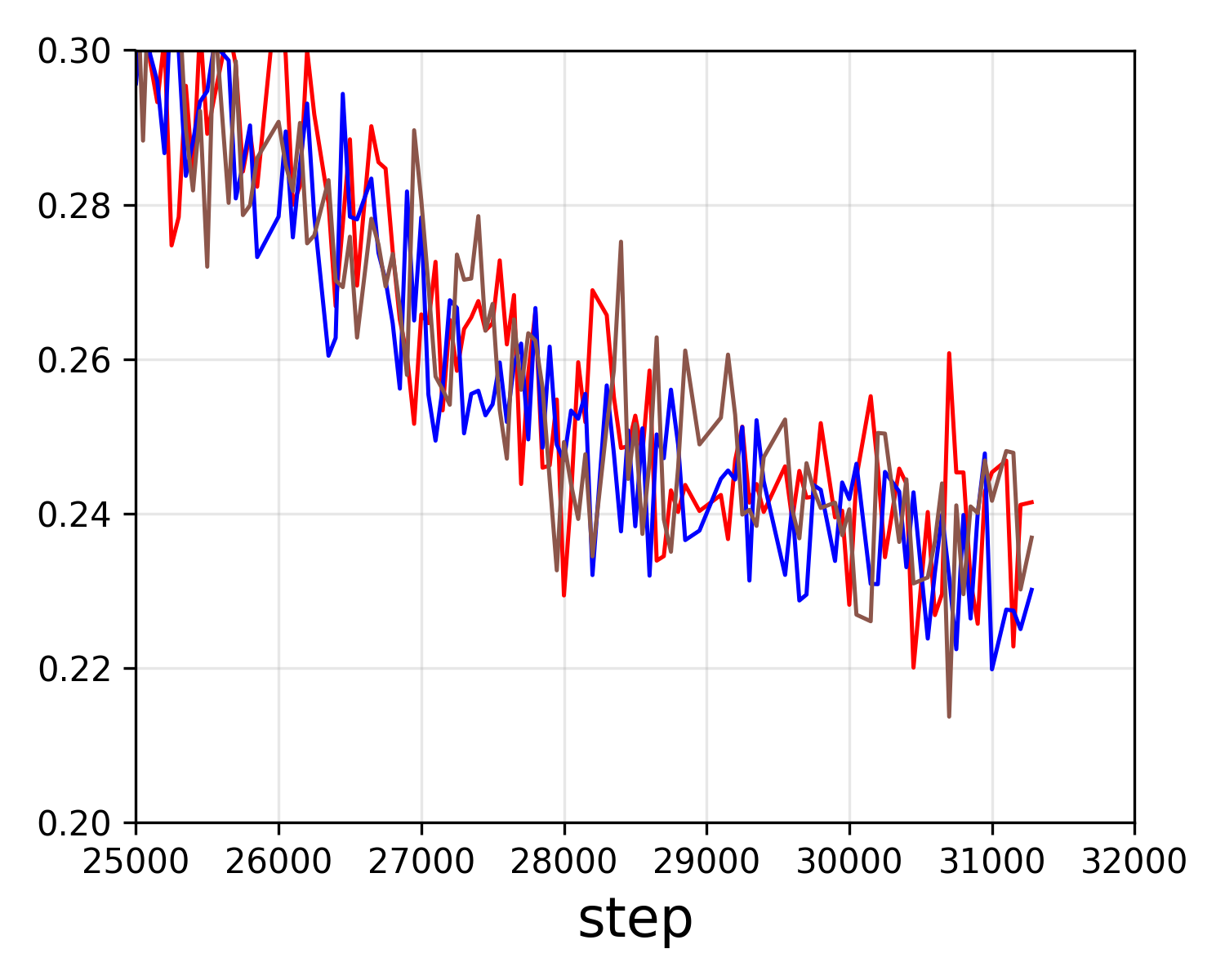}
      \end{minipage}
      \label{fig:dec_contrastive_loss}
      }
    \caption{
    \textbf{Loss visualization with different ViT decoder sizes.} 
    We use \textcolor{red}{M/1}, \textcolor{blue}{B/1} and \textcolor[HTML]{8c564b}{L/1} variants as ViT decoder and draw the loss  curves. As shown in Figures (a) and (b), employing a Large-sized decoder leads to significant training instability. While varying the decoder size has negligible impact on understanding performance, the Base-sized decoder effectively reduces reconstruction loss while maintaining training stability.
        }
    \label{fig:vit_dec}
\end{figure*}

\subsection{VAE latents help unified learning}

A core design of our tokenizer involves feeding VAE latents into a ViT for encoding. 
VAEs are originally designed for low-level visual encoding to capture pixel-level image features.
However, we found that performing unified modeling with VAE latents substantially enhances image generation capabilities without compromising multimodal understanding performance, as shown in~\cref{tab:recon_vae},~\cref{tab:gen_vae}~and~\cref{Tab:wo_vae_und}.
We argue that high-quality VAE latents enable a single ViT encoder to generate a single, unified visual representation that supports both generation and understanding. 
This avoids the need for separate encoding processes followed by concatenation or feature fusion.
Furthermore, within this VAE latent space, we can achieve superior alignment between features for understanding and generation, leading to high performance in both capabilities.

To evaluate the necessity of VAE, we conducted a comparative analysis by training 2 tokenizers with raw image tokens and VAE latents, respectively. 
Specifically, since the downsample rate of FLUX-VAE is $8$, we separately train VAE+ViT-B/2 and ViT-B/16 as the visual tokenizers to keep the same image token number. 
Apart from the difference in the encoder architecture, all other structural components and hyperparameters were kept strictly identical. For both sides, we pretrain at 128$\times$128 resolution for 1000 epochs, and then finetune for 200 epochs at 256$\times$256 for generation, and 224$\times$224 for understanding.

In~\cref{tab:recon_vae}, we report the reconstruction performance with and without VAE. Compared to the standalone ViT, our tokenizer outperforms it by a large margin in rFID (0.216~\vs~0.980), demonstrating our high-quality reconstruction results. Meanwhile, there are marginal gaps in the other three metrics: PSNR, SSIM, and LPIPS. 
Furthermore, we train both tokenizers under RAE framework with 100 epochs to study the effectiveness of VAE in generation. As shown in~\cref{tab:gen_vae}, our model achieves a gFID of 8.45 on ImageNet with the VAE, representing a 1.23 improvement over the 9.68 gFID obtained without it.
In~\cref{Tab:wo_vae_und}, we evaluated the understanding capabilities of the tokenizer when the VAE is removed under both LLaVA-1.5 and LLaVA-NeXT frameworks. The results show a noticeable performance gap compared to~\Model. 
Our tokenizer outperforms the baseline in 4 out of 6 metrics under the LLaVA-1.5 setting; notably, when using the stronger LLaVA-NeXT baseline, our advantage becomes even more pronounced. 
This suggests that although the VAE primarily extracts low-level visual features, it also contributes to the enhancement of multimodal understanding capabilities.
These findings demonstrate that unifying tasks in the VAE latent space yields simultaneous gains in reconstruction fidelity, generative quality, and comprehensive understanding, validating the necessity of the VAE in our unified learning framework.

\subsection{The selection of ViT decoder size}

In the reconstruction branch, we use a ViT and a linear layer as the pixel decoder. 
In~\cref{fig:vit_dec}, we investigate the impact of the size of the ViT decoder on reconstruction and understanding. 
Varying the ViT decoder variants for pixel-level decoding has minimal impact on the understanding branch. As illustrated in~\cref{fig:dec_caption_loss}~and~\cref{fig:dec_contrastive_loss}, the semantic signals remains nearly invariant across different ViT decoder sizes.
Regarding reconstruction learning, increasing the decoder size from M/1 to B/1 results in a significant reduction in pixel-level loss, as shown in~\cref{fig:dec_pixel_loss}. 
However, employing the larger L/1 variant leads to highly unstable training, as depicted by the \textcolor[HTML]{8c564b}{brown} curve. 
In terms of training time, we conduct experiments for 200 epochs for each size. The total training durations for the M/1, B/1, and L/1 decoder on a TPU v4-256 pod are 6.1, 6.0, and 6.9 hours, respectively. 
Taking into account both performance and computational efficiency, employing Base-sized ViT as the decoder strikes an optimal balance.

\subsection{The selection of ViT encoder size}
In~\cref{tab:enc_size}, we explore the scaling behavior of the ViT encoder on the image reconstruction and generation. Generation results are evaluated under RAE pipeline with 100 training epochs.
The results indicate that B-size and L-size tokenizers yield comparable results in reconstruction and generation. In contrast, ~\cref{Tab:understanding}~and~\cref{Tab:und_next}~demonstrate that scaling up to L-size provides substantial gains in multimodal understanding. 
We argue that although increasing the size leads to a more representative tokenizer naturally, the outer VAE architecture imposes a theoretical ceiling on the generative performance.
This bottleneck marginalizes the gains derived from larger encoders, presenting an open problem that requires further investigation in future research.

\begin{table}[t]
    \centering
    \small
        \centering
        \setlength{\tabcolsep}{4pt}
        \caption{
        \textbf{Effectiveness of encoder size in reconstruction and generation.} 
        B size and L size show similar performance in reconstruction and generation ability.
        }
        \begin{tabular}{l|cccc|cccc}
        \toprule
         \multirow{2}{*}{Model} &\multicolumn{4}{c|}{Reconstruction} &\multicolumn{4}{c}{Generation} \\
        & PSNR$\uparrow$ &SSIM$\uparrow$ &LPIPs$\downarrow$ & rFID$\downarrow$ & gFID$\downarrow$  & IS$\uparrow$ & Pre.$\uparrow$ & Rec.$\uparrow$  \\
            \midrule
        
        \Model-B &30.92&0.902&0.053&0.187&8.45&144.7&0.78&0.36 \\
        \Model-L & 30.96&0.902&0.052&0.186&8.89&148.3&0.78&0.36\\
        \bottomrule
        \end{tabular}
        \label{tab:enc_size}
\end{table}

\section{Conclusion}
This work introduces~\Model, a unified vision encoder for both understanding and generation.
We innovatively couple a VAE with a ViT to form a unified architecture, and generate a single, unified representation for different downstream tasks.
For the efficient training of our tokenizer, we propose a new training paradigm with both reconstruction- and semantics-driven signals for joint learning.  Comprehensive evaluations reveal that our model yields superior results across generative and understanding tasks.~\Model~outperforms current other unified tokenizer in reconstruction and generation, and shows competitive ability with CLIP on semantic tasks. 
Furthermore, our architecture enables a mutual promotion relationship between image understanding and generation.
To facilitate future research in the community, we will fully open-source our training code, data, and tokenizer checkpoints.

\bibliographystyle{splncs04}
\bibliography{main}

\begin{thebibliography}{10}
\providecommand{\url}[1]{\texttt{#1}}
\providecommand{\urlprefix}{URL }
\providecommand{\doi}[1]{https://doi.org/#1}

\bibitem{agarwal2025cosmos}
Agarwal, N., Ali, A., Bala, M., Balaji, Y., Barker, E., Cai, T., Chattopadhyay, P., Chen, Y., Cui, Y., Ding, Y., et~al.: Cosmos world foundation model platform for physical ai. arXiv preprint arXiv:2501.03575  (2025)

\bibitem{cherti2023reproducible}
Cherti, M., Beaumont, R., Wightman, R., Wortsman, M., Ilharco, G., Gordon, C., Schuhmann, C., Schmidt, L., Jitsev, J.: Reproducible scaling laws for contrastive language-image learning. In: Proceedings of the IEEE/CVF Conference on Computer Vision and Pattern Recognition. pp. 2818--2829 (2023)

\bibitem{chuang2025meta}
Chuang, Y.S., Li, Y., Wang, D., Yeh, C.F., Lyu, K., Raghavendra, R., Glass, J., Huang, L., Weston, J., Zettlemoyer, L., et~al.: Meta clip 2: A worldwide scaling recipe. arXiv preprint arXiv:2507.22062  (2025)

\bibitem{comanici2025gemini}
Comanici, G., Bieber, E., Schaekermann, M., Pasupat, I., Sachdeva, N., Dhillon, I., Blistein, M., Ram, O., Zhang, D., Rosen, E., et~al.: Gemini 2.5: Pushing the frontier with advanced reasoning, multimodality, long context, and next generation agentic capabilities. arXiv preprint arXiv:2507.06261  (2025)

\bibitem{cui2025emu3}
Cui, Y., Chen, H., Deng, H., Huang, X., Li, X., Liu, J., Liu, Y., Luo, Z., Wang, J., Wang, W., et~al.: Emu3. 5: Native multimodal models are world learners. arXiv preprint arXiv:2510.26583  (2025)

\bibitem{deng2025emerging}
Deng, C., Zhu, D., Li, K., Gou, C., Li, F., Wang, Z., Zhong, S., Yu, W., Nie, X., Song, Z., et~al.: Emerging properties in unified multimodal pretraining. arXiv preprint arXiv:2505.14683  (2025)

\bibitem{deng2009imagenet}
Deng, J., Dong, W., Socher, R., Li, L.J., Li, K., Fei-Fei, L.: Imagenet: A large-scale hierarchical image database. In: 2009 IEEE conference on computer vision and pattern recognition. pp. 248--255. Ieee (2009)

\bibitem{esser2024scaling}
Esser, P., Kulal, S., Blattmann, A., Entezari, R., M{\"u}ller, J., Saini, H., Levi, Y., Lorenz, D., Sauer, A., Boesel, F., et~al.: Scaling rectified flow transformers for high-resolution image synthesis. In: Forty-first international conference on machine learning (2024)

\bibitem{esser2021taming}
Esser, P., Rombach, R., Ommer, B.: Taming transformers for high-resolution image synthesis. In: Proceedings of the IEEE/CVF conference on computer vision and pattern recognition. pp. 12873--12883 (2021)

\bibitem{fan2025unified}
Fan, L., Tang, L., Qin, S., Li, T., Yang, X., Qiao, S., Steiner, A., Sun, C., Li, Y., Zhu, T., et~al.: Unified autoregressive visual generation and understanding with continuous tokens. arXiv preprint arXiv:2503.13436  (2025)

\bibitem{fang2023data}
Fang, A., Jose, A.M., Jain, A., Schmidt, L., Toshev, A., Shankar, V.: Data filtering networks. arXiv preprint arXiv:2309.17425  (2023)

\bibitem{mme}
Fu, C., Chen, P., Shen, Y., Qin, Y., Zhang, M., Lin, X., Yang, J., Zheng, X., Li, K., Sun, X., et~al.: Mme: A comprehensive evaluation benchmark for multimodal large language models. arXiv preprint arXiv:2306.13394  (2023)

\bibitem{gadre2023datacomp}
Gadre, S.Y., Ilharco, G., Fang, A., Hayase, J., Smyrnis, G., Nguyen, T., Marten, R., Wortsman, M., Ghosh, D., Zhang, J., et~al.: Datacomp: In search of the next generation of multimodal datasets. Advances in Neural Information Processing Systems  \textbf{36},  27092--27112 (2023)

\bibitem{heinrich2025radiov2}
Heinrich, G., Ranzinger, M., Yin, H., Lu, Y., Kautz, J., Tao, A., Catanzaro, B., Molchanov, P.: Radiov2. 5: Improved baselines for agglomerative vision foundation models. In: Proceedings of the Computer Vision and Pattern Recognition Conference. pp. 22487--22497 (2025)

\bibitem{gqa}
Hudson, D.A., Manning, C.D.: Gqa: A new dataset for real-world visual reasoning and compositional question answering. In: CVPR (2019)

\bibitem{huh2024platonic}
Huh, M., Cheung, B., Wang, T., Isola, P.: The platonic representation hypothesis. arXiv preprint arXiv:2405.07987  (2024)

\bibitem{hurst2024gpt}
Hurst, A., Lerer, A., Goucher, A.P., Perelman, A., Ramesh, A., Clark, A., Ostrow, A., Welihinda, A., Hayes, A., Radford, A., et~al.: Gpt-4o system card. arXiv preprint arXiv:2410.21276  (2024)

\bibitem{flux2024}
Labs, B.F.: Flux. \url{https://github.com/black-forest-labs/flux} (2024)

\bibitem{labs2025flux}
Labs, B.F., Batifol, S., Blattmann, A., Boesel, F., Consul, S., Diagne, C., Dockhorn, T., English, J., English, Z., Esser, P., et~al.: Flux. 1 kontext: Flow matching for in-context image generation and editing in latent space. arXiv preprint arXiv:2506.15742  (2025)

\bibitem{leng2025repa}
Leng, X., Singh, J., Hou, Y., Xing, Z., Xie, S., Zheng, L.: Repa-e: Unlocking vae for end-to-end tuning with latent diffusion transformers. arXiv preprint arXiv:2504.10483  (2025)

\bibitem{seedbench}
Li, B., Ge, Y., Ge, Y., Wang, G., Wang, R., Zhang, R., Shan, Y.: Seed-bench: Benchmarking multimodal large language models. In: CVPR (2024)

\bibitem{li2025openvision}
Li, X., Liu, Y., Tu, H., Xie, C.: Openvision: A fully-open, cost-effective family of advanced vision encoders for multimodal learning. In: Proceedings of the IEEE/CVF International Conference on Computer Vision. pp. 3977--3987 (2025)

\bibitem{li2024if}
Li, X., Tu, H., Hui, M., Wang, Z., Zhao, B., Xiao, J., Ren, S., Mei, J., Liu, Q., Zheng, H., et~al.: What if we recaption billions of web images with llama-3? arXiv preprint arXiv:2406.08478  (2024)

\bibitem{li2023clipa}
Li, X., Wang, Z., Xie, C.: Clipa-v2: Scaling clip training with 81.1\% zero-shot imagenet accuracy within a \$10,000 budget; an extra \$4,000 unlocks 81.8\% accuracy. arXiv preprint arXiv:2306.15658  (2023)

\bibitem{li2023inverse}
Li, X., Wang, Z., Xie, C.: An inverse scaling law for clip training. Advances in Neural Information Processing Systems  \textbf{36},  49068--49087 (2023)

\bibitem{pope}
Li, Y., Du, Y., Zhou, K., Wang, J., Zhao, X., Wen, J.R.: Evaluating object hallucination in large vision-language models. In: EMNLP (2023)

\bibitem{liao2025mogao}
Liao, C., Liu, L., Wang, X., Luo, Z., Zhang, X., Zhao, W., Wu, J., Li, L., Tian, Z., Huang, W.: Mogao: An omni foundation model for interleaved multi-modal generation. arXiv preprint arXiv:2505.05472  (2025)

\bibitem{lin2025uniworld}
Lin, B., Li, Z., Cheng, X., Niu, Y., Ye, Y., He, X., Yuan, S., Yu, W., Wang, S., Ge, Y., et~al.: Uniworld: High-resolution semantic encoders for unified visual understanding and generation. arXiv preprint arXiv:2506.03147  (2025)

\bibitem{lin2014microsoft}
Lin, T.Y., Maire, M., Belongie, S., Hays, J., Perona, P., Ramanan, D., Doll{\'a}r, P., Zitnick, C.L.: Microsoft coco: Common objects in context. In: European conference on computer vision. pp. 740--755. Springer (2014)

\bibitem{liu2024improved}
Liu, H., Li, C., Li, Y., Lee, Y.J.: Improved baselines with visual instruction tuning. In: Proceedings of the IEEE/CVF conference on computer vision and pattern recognition. pp. 26296--26306 (2024)

\bibitem{liu2024clips}
Liu, Y., Li, X., Wang, Z., Zhao, B., Xie, C.: Clips: An enhanced clip framework for learning with synthetic captions. arXiv preprint arXiv:2411.16828  (2024)

\bibitem{liu2025openvision}
Liu, Y., Li, X., Zhang, L., Wang, Z., Zheng, Z., Zhou, Y., Xie, C.: Openvision 2: A family of generative pretrained visual encoders for multimodal learning. arXiv preprint arXiv:2509.01644  (2025)

\bibitem{liu2025tuna}
Liu, Z., Ren, W., Liu, H., Zhou, Z., Chen, S., Qiu, H., Huang, X., An, Z., Yang, F., Patel, A., et~al.: Tuna: Taming unified visual representations for native unified multimodal models. arXiv preprint arXiv:2512.02014  (2025)

\bibitem{ma2025unitok}
Ma, C., Jiang, Y., Wu, J., Yang, J., Yu, X., Yuan, Z., Peng, B., Qi, X.: Unitok: A unified tokenizer for visual generation and understanding. arXiv preprint arXiv:2502.20321  (2025)

\bibitem{ma2024sit}
Ma, N., Goldstein, M., Albergo, M.S., Boffi, N.M., Vanden-Eijnden, E., Xie, S.: Sit: Exploring flow and diffusion-based generative models with scalable interpolant transformers. In: European Conference on Computer Vision. pp. 23--40. Springer (2024)

\bibitem{peebles2023scalable}
Peebles, W., Xie, S.: Scalable diffusion models with transformers. In: Proceedings of the IEEE/CVF international conference on computer vision. pp. 4195--4205 (2023)

\bibitem{qu2025tokenflow}
Qu, L., Zhang, H., Liu, Y., Wang, X., Jiang, Y., Gao, Y., Ye, H., Du, D.K., Yuan, Z., Wu, X.: Tokenflow: Unified image tokenizer for multimodal understanding and generation. In: Proceedings of the Computer Vision and Pattern Recognition Conference. pp. 2545--2555 (2025)

\bibitem{clip}
Radford, A., Kim, J.W., Hallacy, C., Ramesh, A., Goh, G., Agarwal, S., Sastry, G., Askell, A., Mishkin, P., Clark, J., et~al.: Learning transferable visual models from natural language supervision. In: ICML (2021)

\bibitem{ranzinger2024radio}
Ranzinger, M., Heinrich, G., Kautz, J., Molchanov, P.: Am-radio: Agglomerative vision foundation model reduce all domains into one. In: Proceedings of the IEEE/CVF conference on computer vision and pattern recognition. pp. 12490--12500 (2024)

\bibitem{rombach2022high}
Rombach, R., Blattmann, A., Lorenz, D., Esser, P., Ommer, B.: High-resolution image synthesis with latent diffusion models. In: Proceedings of the IEEE/CVF conference on computer vision and pattern recognition. pp. 10684--10695 (2022)

\bibitem{scienceqa}
Saikh, T., Ghosal, T., Mittal, A., Ekbal, A., Bhattacharyya, P.: Scienceqa: A novel resource for question answering on scholarly articles. In: IJDL (2022)

\bibitem{schuhmann2022laion}
Schuhmann, C., Beaumont, R., Vencu, R., Gordon, C., Wightman, R., Cherti, M., Coombes, T., Katta, A., Mullis, C., Wortsman, M., et~al.: Laion-5b: An open large-scale dataset for training next generation image-text models. Advances in neural information processing systems  \textbf{35},  25278--25294 (2022)

\bibitem{tang2025unilip}
Tang, H., Xie, C., Bao, X., Weng, T., Li, P., Zheng, Y., Wang, L.: Unilip: Adapting clip for unified multimodal understanding, generation and editing. arXiv preprint arXiv:2507.23278  (2025)

\bibitem{tschannen2025siglip}
Tschannen, M., Gritsenko, A., Wang, X., Naeem, M.F., Alabdulmohsin, I., Parthasarathy, N., Evans, T., Beyer, L., Xia, Y., Mustafa, B., et~al.: Siglip 2: Multilingual vision-language encoders with improved semantic understanding, localization, and dense features. arXiv preprint arXiv:2502.14786  (2025)

\bibitem{wan2025wan}
Wan, T., Wang, A., Ai, B., Wen, B., Mao, C., Xie, C.W., Chen, D., Yu, F., Zhao, H., Yang, J., et~al.: Wan: Open and advanced large-scale video generative models. arXiv preprint arXiv:2503.20314  (2025)

\bibitem{wang2024omnitokenizer}
Wang, J., Jiang, Y., Yuan, Z., Peng, B., Wu, Z., Jiang, Y.G.: Omnitokenizer: A joint image-video tokenizer for visual generation. Advances in Neural Information Processing Systems  \textbf{37},  28281--28295 (2024)

\bibitem{wu2024vila}
Wu, Y., Zhang, Z., Chen, J., Tang, H., Li, D., Fang, Y., Zhu, L., Xie, E., Yin, H., Yi, L., et~al.: Vila-u: a unified foundation model integrating visual understanding and generation. arXiv preprint arXiv:2409.04429  (2024)

\bibitem{xie2025show}
Xie, J., Yang, Z., Shou, M.Z.: Show-o2: Improved native unified multimodal models. arXiv preprint arXiv:2506.15564  (2025)

\bibitem{xu2023demystifying}
Xu, H., Xie, S., Tan, X.E., Huang, P.Y., Howes, R., Sharma, V., Li, S.W., Ghosh, G., Zettlemoyer, L., Feichtenhofer, C.: Demystifying clip data. arXiv preprint arXiv:2309.16671  (2023)

\bibitem{yao2025reconstruction}
Yao, J., Yang, B., Wang, X.: Reconstruction vs. generation: Taming optimization dilemma in latent diffusion models. In: Proceedings of the Computer Vision and Pattern Recognition Conference. pp. 15703--15712 (2025)

\bibitem{yu2022coca}
Yu, J., Wang, Z., Vasudevan, V., Yeung, L., Seyedhosseini, M., Wu, Y.: Coca: Contrastive captioners are image-text foundation models. arXiv preprint arXiv:2205.01917  (2022)

\bibitem{yu2024representation}
Yu, S., Kwak, S., Jang, H., Jeong, J., Huang, J., Shin, J., Xie, S.: Representation alignment for generation: Training diffusion transformers is easier than you think. arXiv preprint arXiv:2410.06940  (2024)

\bibitem{zhai2023sigmoid}
Zhai, X., Mustafa, B., Kolesnikov, A., Beyer, L.: Sigmoid loss for language image pre-training. In: Proceedings of the IEEE/CVF international conference on computer vision. pp. 11975--11986 (2023)

\bibitem{zheng2025diffusion}
Zheng, B., Ma, N., Tong, S., Xie, S.: Diffusion transformers with representation autoencoders. arXiv preprint arXiv:2510.11690  (2025)

\end{thebibliography}

\end{document}